\documentclass[useAMS,usenatbib]{mn2e}
\usepackage{amsmath,astrojournals,fleqn,graphicx,txfonts}
\newcommand  {\B}[1]{{\bmath{#1}}}
\newcommand  {\s}[1]{{\mathsf{#1}}}

\newcommand  {\divv}{{\B{\nabla}{\cdot}\B{\upsilon}}}
\newcommand  {\curlv}{{\B{\nabla}{\,\times\,}\B{\upsilon}}}
\newcommand  {\diva}{{\B{\nabla}{\cdot}\dot{\B{\upsilon}}}}
\newcommand  {\Dtdivv}{{\dot{\B{\nabla}}{\cdot}\B{\upsilon}}}
\title{Inviscid SPH}
\author[Lee~Cullen \& Walter~Dehnen]
 {Lee Cullen%
\thanks{Email: lee.cullen@astro.le.ac.uk, walter.dehnen@astro.le.ac.uk}
  and Walter Dehnen$^{\star}$\\
  Department of Physics \& Astronomy,
  University of Leicester,
  Leicester, LE1~7RH
}
\date{Accepted .
      Received ;
      }
\pagerange{\pageref{firstpage}--\pageref{lastpage}}
\pubyear{2009}
\begin{document}
\maketitle
\label{firstpage}
\begin{abstract}
  In smooth-particle hydrodynamics (SPH), artificial viscosity is necessary
  for the correct treatment of shocks, but often generates unwanted
  dissipation away from shocks. We present a novel method of controlling the
  amount of artificial viscosity, which uses the total time derivative of the
  velocity divergence as shock indicator and aims at completely eliminating
  viscosity away from shocks. We subject the new scheme to numerous tests and
  find that the method works at least as well as any previous technique in the
  strong-shock regime, but becomes virtually inviscid away from shocks, while
  still maintaining particle order. In particular sound waves or oscillations
  of gas spheres are hardly damped over many periods.
\end{abstract}

\section{Introduction}
\label{sec:intro}
Smooth-particle hydrodynamics (SPH) is a Lagrangian method for modelling
fluid dynamics, pioneered by \cite{GingoldMonaghan1977} and
\cite{Lucy1977}. Instead of discretising the fluid quantities, such as
density, velocity, and temperature, on a fixed grid as in Eulerian methods,
the fluid is represented by a discrete set of moving particles acting as
interpolation points.  Due to its Lagrangian nature, SPH models regions of
higher density with higher resolution with the ability to simulate large
dynamic ranges. This makes it particularly useful in astrophysics, where it is
used to model galaxy and star formation, stellar collisions, and accretion
discs.

The core of SPH is the kernel estimator: the fluid density is \emph{estimated}
from the masses $m_i$ and positions $\B{x}_i$ of the particles via%
\footnote{We use the symbol $\,\hat{}\,$ to denote a local \emph{estimate} --
  in many SPH-related publications the distinction between actual and
  estimated quantities is not clearly made, confusing the discussion.}
\begin{equation}
  \label{eq:est:rho} \textstyle
  \hat{\rho}(\B{x}_i) = \sum_j m_j\,W(|\B{x}_i-\B{x_j}|,h_i),
\end{equation}
where $W$ is the kernel function and $h_i$ the SPH smoothing length%
\footnote{In this study we use the convention that the kernel has finite
  support of one smoothing length radius, i.e.\ $W=0$ for
  $|\B{x}_i-\B{x}_j|>h$.} for the $i$th particle. Similar estimates
$\hat{F}(\B{x})$ for the value of any field $F(\B{x})$ can be obtained from
its discretised values $F_i$. By applying these estimators to the fluid
equations governing mass, momentum and energy, discrete equations for the SPH
particle positions $\B{x}_i$ and other properties (such as internal energy
$u_i$) can be obtained. Together with an appropriate time integration method,
these constitute a concrete SPH scheme.

Unfortunately, this process is not unique and since its inception the SPH
method has undergone many refinements such as individual particle smoothing
lengths and viscosities, as well as many alternative derivations of the SPH
equations, leading to a plethora of SPH methods. While formally these various
schemes differ only in their error terms, their conservation and stability
properties can be quite different. This has lead to the unfortunate situation
that the shortcomings of a few such implementations are often blamed on the
general SPH concept per se.

However, \cite{SpringelHernquist2002} have pointed out that SPH equations
derived from a variational principle are not only unique, but also
conservative. Such SPH equations are most simply obtained as the
Euler-Lagrange equations derived from an SPH Lagrangian $\mathcal{L}$
representing the Lagrangian of the fluid system. Once $\mathcal{L}$ is chosen,
the SPH equations follow uniquely (see Appendix~\ref{app:pressure} for a
typical example). Complementing these with a symplectic integrator, such as
the standard leap-frog, results in a SPH scheme which by construction
conserves the total mass, momentum, angular momentum, energy, and entropy.

\begin{figure}
  \centerline{
    \resizebox{70mm}{!}{\includegraphics{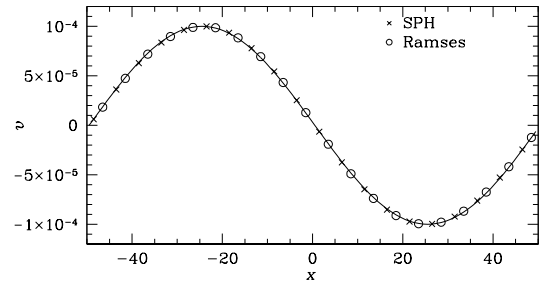}} }
  \caption{
    \label{fig:sound_wave}
    A 1D sinusoidal sound wave with velocity amplitude $10^{-4}c$ and
    $\gamma=1.4$ propagated for 50 periods with SPH without artificial
    viscosity using 100 particles and with a grid code
    \citep[\textsf{Ramses},][]{Teyssier2002} using 128 cells (only every fifth
    particle or grid cell is plotted). Both methods preserve the wave
    amplitude and period, demonstrating their dissipation-less nature.}
\end{figure} 
The conservation of entropy means that SPH is \emph{dissipation-less}, as
demonstrated in Fig.~\ref{fig:sound_wave}. In real fluids, however, entropy
increases in shocks, where particle collisions randomise their velocities
generating heat and entropy. This basic collisional mechanism is inherent to
all fluids (except for dust and collisionless plasma, which therefore may not
be considered fluids) and prevents the flow from becoming multi-valued. In SPH
\emph{artificial} viscosity is needed to dissipate local velocity differences
and convert them into heat, which generates entropy and prevents
inter-penetration of SPH particles and thus a multi-valued flow.

Since the artificial viscosity required for this goal is usually much stronger
than the actual physical viscosity, it also causes unphysical dissipation away
from shocks. While it may be possible for certain simulations to select the
magnitude of the viscosity to minimise such undesired dissipation, in general
the adverse effect of artificial viscosity is unknown prior to any simulation
and, possibly, even afterwards. For example, when simulating the effect of a
perturbing massive body on a pulsating star, it may be very difficult to
distinguish this effect from that induced by artificial viscosity. Another
example is the case of a differentially rotating disc, where artificial
viscosity causes spurious angular momentum transport.

Since viscosity is a dissipative process, the corresponding SPH equations
cannot be derived from a variational principle, and we are back to ad-hoc
methods for deriving them. Most SPH simulations to date still use a rather
simple artificial viscosity, which effectively amounts to modelling a viscous
fluid and quickly damps away any oscillations, such as sound waves or stellar
pulsations, and impedes shear flows. While suggestions have been made to
\emph{reduce} such unwanted dissipation, our goal here is to \emph{eliminate}
it. To this end we introduce a novel method of controlling the amount of
artificial viscosity, such that away from shocks the modelled flow is
virtually inviscid.

Section~\ref{sec:AV} describes SPH artificial viscosity and previous efforts
to reduce its adverse effects, while our new method is outlined in
Section~\ref{sec:AV:novel}. The ability of the new scheme to reduce artificial
viscosity but also to capture shocks is demonstrated in Sections
~\ref{sec:test:suppress} and \ref{sec:test:shock}, respectively. Finally,
Section~\ref{sec:summ} concludes our study.

\section{Reducing unwanted artificial viscosity} \label{sec:AV}

\subsection{Standard SPH artificial viscosity} \label{sec:AV:SPH}
The traditional form of artificial viscosity \citep[e.g.][]{Monaghan1992} adds
the following terms to the momentum and energy equations, allowing the
conversion of kinetic energy into heat.
\begin{subequations}
  \begin{eqnarray}
    \label{eq:dv}
    \left(\dot{\B{\upsilon}}_i\right)_{\mathrm{AV}} &=& \textstyle
    - \sum_j m_j\,\Pi_{i\!j}\,\B{\nabla}_i \overline{W}_{i\!j} \\
  \label{eq:du}
    \left(\dot{u}_i\right)_{\mathrm{AV}} &=& \textstyle \frac{1}{2} \sum_j
    m_j\,\Pi_{i\!j} \;\B{\upsilon}_{i\!j} \cdot \B{\nabla}_i
    \overline{W}_{i\!j}
  \end{eqnarray}
\end{subequations}
with the average kernel
\begin{equation} 
  \label{eq:Wbar}
  \overline{W}_{i\!j} = \tfrac{1}{2}\left(W(|\B{x}_{i\!j}|,h_i) +
    W(|\B{x}_{i\!j}|,h_j)\right).
\end{equation}
Here, $\B{x}_{i\!j}\equiv\B{x}_i-\B{x}_j$ and
$\B{\upsilon}_{i\!j}\equiv\B{\upsilon}_i-\B{\upsilon}_j$, while $h_i$ is the
individual adaptive smoothing length of each SPH particle (for details on how
$h_i$ is adapted see Appendix~\ref{sec:adapt}).  The artificial viscosity term
is given by \citep{GingoldMonaghan1983}
\begin{equation}
  \label{eq:Pi}
  \Pi_{i\!j} = \begin{cases}
    \displaystyle
    \frac{-\alpha\,\bar{c}_{i\!j}\,\mu_{i\!j} + \beta\,\mu^2_{i\!j}} 
         {\bar{\hat{\rho}}_{i\!j}} & 
         \text{for $\B{\upsilon}_{i\!j}\cdot \B{x}_{i\!j} < 0$} \\
    0 &  \text{otherwise}
  \end{cases}
\end{equation}
with
\begin{equation}
\label{eq:mu}
\mu_{i\!j} = \frac{\bar{h}_{i\!j}\,\B{\upsilon}_{i\!j}
  \cdot \B{x}_{i\!j}} {\B{x}_{i\!j}^2 +
  \epsilon^2}
\end{equation}
($\bar{h}_{i\!j}\equiv[h_i+h_j]/2$ and likewise for the average sound speed
$\bar{c}_{i\!j}$ and estimated density $\bar{\hat{\rho}}_{i\!j}$). Since
$\Pi_{i\!j}\,{=}\,0$ for receding particle pairs, artificial viscosity does
not affect expanding flows. This functional form of SPH artificial viscosity
may seem rather ad-hoc, but it is reasonably well motivated and emerged as the
most useful one amongst several methods \citep{GingoldMonaghan1983}. Moreover,
it is equivalent to the form of dissipation implicit in Riemann solvers
\citep{Monaghan1997}.

By expanding density and velocity in a Taylor series around $\B{x}_i$, it is
straightforward to show that these terms correspond to both a shear and a bulk
viscosity. More quantitatively, if one assumes that, other than in
equation~(\ref{eq:Pi}), artificial viscosity acts between approaching and
receding neighbours and that $\beta=0$, the corresponding shear and bulk
viscosity coefficients are \citep[e.g.][]{MeglickiEtal1993}
$\eta\,{=}\,\frac{1}{2} \alpha \kappa h c \rho$ and $\zeta\,{=}\,\frac{5}{3}
\eta$, respectively, where the factor $\kappa$ is of order unity and depends
on the functional form of the kernel. This implies that artificial viscosity
decreases with increasing resolution (smaller $h$). Thus, a straightforward
though expensive way to reduce unwanted dissipation is to increase the
resolution. In fact, one motivation for reducing artificial viscosity is to
avoid this purely numerical necessity for high resolution.

Most SPH applications to date use the above treatment with $\alpha=1$. The
widely used code \textsc{gadget-2} \citep{Springel2005} employs a fixed
$\alpha$ chosen at the start of the simulation (though
\citeauthor{DolagEtAl2005}, \citeyear{DolagEtAl2005}, have implemented into
\textsc{gadget-2} the improved method described in \S\ref{sec:AV:MM}
below). Clearly, in complex situations, where strong and weak shocks are
present as well as converging flows, any choice for $\alpha$ is
unsatisfactory, leading to bad treatment of strong shocks, over-damping of
converging flows, or both.

\subsection{Balsara's method} \label{sec:AV:Balsara}
The purpose of artificial viscosity is to allow for entropy generation across
shocks and to stop particle interpenetration. To this end, only bulk viscosity
is required, but the inherent shear viscosity is unnecessary. What is worse,
this shear viscosity may seriously compromise simulations of shear flows, such
as in a differentially rotating gas disc. In an effort to reduce the resulting
artificial angular momentum dissipation, \cite{Balsara1995} proposed to
multiply $\Pi_{i\!j}$ with a reduction factor $\bar{f}_{i\!j}=(f_i+f_j)/2$ with
\begin{equation} \label{eq:balsara}
  f_i = \frac{\left|\divv_i \right|}
  {\left| \divv_i \right|
    + \left|\curlv_i \right| }
\end{equation}
(with velocity divergence and curl estimated using the SPH kernel estimator).
This term diminishes the effect of artificial viscosity whenever the vorticity
dominates the convergence. However, this method only reduces (but does not
eliminate) unwanted dissipation in the presence of a rotating shear flow.

\begin{figure}
  \centerline{
    \resizebox{75mm}{!}{\includegraphics{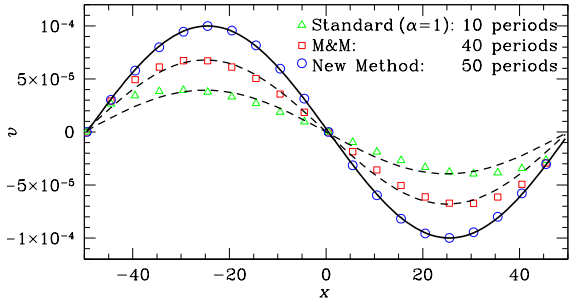}}
  }
  \caption{As Fig.~\ref{fig:sound_wave}, but for SPH with standard
    ($\alpha=1)$ or 
    Morris \& Monaghan (1997)
    artificial viscosity, as well as our new method (only every fifth particle
    is plotted). Also shown are the undamped wave (\emph{solid}) and
    lower-amplitude sinusoidals (\emph{dashed}). Only with our method the wave
    propagates undamped, very much like SPH without any viscosity, as in
    Fig.~\ref{fig:sound_wave}. \label{fig:damp_sound} }
\end{figure}
\subsection{The method of \citeauthor{MorrisMonaghan1997}}
\label{sec:AV:MM}
Standard SPH artificial viscosity acts whenever the flow of the fluid
converges, even if only weakly. For example, when a pulsating star contracts
artificial viscosity damps its pulsation. Exactly the same happens to ordinary
sound waves: standard SPH viscosity damps them, as demonstrated in
Fig.~\ref{fig:damp_sound}, the faster the shorter the wave length (because
these are more poorly resolved).

With this in mind, \cite{MorrisMonaghan1997} proposed to adapt the strength of
artificial viscosity to the local convergence of the flow. To this end, they
introduced the concept of individual adaptive viscosities $\alpha_i$ for each
particle, replaced $\alpha$ in equation~(\ref{eq:Pi}) by
$\bar{\alpha}_{i\!j}=(\alpha_i+\alpha_j)/2$, and set
$\beta\propto\bar{\alpha}_{i\!j}$. The individual viscosities are adapted
according to the differential equation
\begin{equation} 
  \label{eq:davdt}
  \dot{\alpha}_i = 
  (\alpha_{\min}-\alpha_i)/\tau_i
  + S_{\!i}
\end{equation}
with the velocity-based source term
\begin{equation} 
  \label{eq:S:MM}
  S_{\!i} = \max\big\{-\divv_i,\,0\big\}.
\end{equation}
and the decay time\footnote{The factor 2 in the denominator of
  equation~(\ref{eq:tau}) accounts for the difference in the definition of the
  smoothing length $h$ between us and \cite{MorrisMonaghan1997}.}
\begin{equation} \label{eq:tau}
  \tau_i = h_i/(2 \ell c_i).
\end{equation}
Here, $\alpha_{\min}=0.1$ constitutes a lower limit for the artificial
viscosity such that $\alpha_i=\alpha_{\min}$ for non-convergent flows. For a
convergent flow, on the other hand, $\alpha_i$ grows above that value,
guaranteeing the proper treatment of shocks. In the post-shock region, the
flow is no longer convergent and $\alpha_i$ decays back to $\alpha_{\min}$ on
the time scale $\tau_i$ (typically $\ell=0.1-0.2$). This method reduces the
artificial viscosity away from shocks by an order of magnitude compared to
standard SPH and gives equally accurate post and pre-shock solutions
\citep{MorrisMonaghan1997}.

More recently, \cite*{RosswogEtAl2000} proposed to alter the adaption equation
(\ref{eq:davdt}) to%
\footnote{ This is equivalent to keeping (\ref{eq:davdt})
  but multiplying the source term (\ref{eq:S:MM}) by $(\alpha_{\max}-\alpha)$,
  which is what \citeauthor{RosswogEtAl2000} actually did.}%
\begin{equation} 
  \label{eq:davdt:alt}
  \dot{\alpha}_i = (\alpha_{\min}-\alpha_i)/\tau_i
  + (\alpha_{\max}-\alpha_i)\,S_{\!i}
\end{equation}
with $\alpha_{\max}=1.5$, while \cite{Price2004} advocated
$\alpha_{\max}=2$. The effect of this alteration is first to prevent
$\alpha_i$ to exceed $\alpha_{\max}$ and second to increase $\dot{\alpha}_i$
for small $\alpha_i$, which ensures a faster viscosity growth, resulting in
somewhat better treatment of shocks \citep{Price2004}. This method may also be
combined with the Balsara switch by applying the reduction factor
(\ref{eq:balsara}) either to $\Pi_{i\!j}$ \citep{RosswogEtAl2000} or to
$S_{\!i}$ \citep{MorrisMonaghan1997,WetzsteinEtAl2009}.

The scheme of equations (\ref{eq:S:MM}), (\ref{eq:tau}) and
(\ref{eq:davdt:alt}) with $\alpha_{\min}=0.1$, $\alpha_{\max}=2$ and
$\ell=0.1$ is the current state of the art for SPH and is implemented in the
codes \textsc{phantom} (by Daniel Price) and \textsc{vine}
\citep{WetzsteinEtAl2009}. In sections \ref{sec:test:suppress} and
\ref{sec:test:shock}, we will frequently compare our novel scheme (to be
described below) with this method and refer to it as the `M\&M method' or the
`\cite{Price2004} version of the M\&M method' as opposed to the `original M\&M
method', which uses equation (\ref{eq:davdt}) instead of (\ref{eq:davdt:alt}).

\subsection{Critique of the M\&M method} \label{sec:MM:crit}
The M\&M method certainly constitutes a large improvement over standard SPH,
but low-viscosity flows, typical for many astrophysical fluids, are still
inadequately modelled. After studying this and related methods in detail,
we identify the following problems.

First, any $\alpha_{\min}{\,>\,}0$ results in unwanted dissipation, for
example of sound waves (see Fig.~\ref{fig:damp_sound}) or stellar pulsations
(see \S\ref{sec:test:poly}), yet the M\&M method requires
$\alpha_{\min}{\,\approx\,}0.1$. This necessity has been established by
numerous tests \citep[most notably of ][]{Price2004} and is understood to
originate from the requirement to `maintain order amongst the particles away
from shocks' \citep{MorrisMonaghan1997}.

Second, there is a delay between the peak in the viscosity $\alpha$ and the
shock front (see Fig.~\ref{fig:source}): the particle viscosities are still
rising when the shock arrives. One reason for this lag is that integrating the
differential equation~(\ref{eq:davdt:alt}) increases $\alpha_i$ too slowly:
the asymptotic value
\begin{equation} \label{eq:a:asym}
  \alpha_{\mathrm{s}} = \frac{\alpha_{\min}+\alpha_{\max}\,S_{\!i}\tau_i}
        {1+S_{\!i}\tau_i}
\end{equation}
is hardly ever reached before the shock arrives (and $S_{\!i}$ decreases).

Third, the source term (\ref{eq:S:MM}) does not distinguish between pre- and
post-shock regions: for a symmetrically smoothed shock it peaks at the exact
shock position \citep[in practice the peak occurs one particle separation in
  front of the shock,][see also
  Fig.~\ref{fig:source}]{MorrisMonaghan1997}. However, immediately behind the
shock (or more precisely the minimum of $\divv$), the (smoothed) flow is still
converging and hence $\alpha$ continues to increase without need. A further
problem is the inability of the source term (\ref{eq:S:MM}) to distinguish
between velocity discontinuities and convergent flows.

Finally, in strong shear flows the estimation of the velocity divergence
$\divv$, needed in (\ref{eq:S:MM}), often suffers from substantial errors (see
Appendix~\ref{app:divv:fail} for the reason), driving artificial viscosity
without need. This especially compromises simulations of differentially
rotating discs even when using the Balsara switch.

\section{A novel artificial viscosity scheme} \label{sec:AV:novel}
Our aim is a method which overcomes all the issues identified in
\S\ref{sec:MM:crit} above and in particular gives $\alpha_i\to0$ away from
shocks.  To this end, we introduce a new shock indicator in \S\ref{sec:AV:A},
a novel technique for adapting $\alpha_i$ in \S\ref{sec:AV:new}, and a method
to suppress false compression detections due to the presence of strong shear
in \S\ref{sec:shear}.

\subsection{A novel shock indicator} \label{sec:AV:A}
We need a shock indicator which not only distinguishes shocks from convergent
flows, but, unlike $\divv$, also discriminates between pre- and post-shock
regions. This requires (at least) a second-order derivative of the flow
velocity and we found the total time derivative of the velocity divergence,
$\Dtdivv \equiv \mathrm{d}(\divv)/\mathrm{d} t$, to be most useful. As is
evident from differentiating the continuity equation,
\begin{equation}
  -\Dtdivv = \mathrm{d}^2\ln\rho/\mathrm{d}t^2,
\end{equation}
$\Dtdivv<0$ indicates an non-linear density increase and a steepening of the
flow convergence, as is typical for any pre-shock region.  Conversely, in the
post-shock region $\Dtdivv>0$. This suggests to consider only negative values
and, in analogy with equation (\ref{eq:S:MM}), we define the new shock
indicator
\begin{equation} \label{eq:A}
  A_i = \xi_i\,\max\big\{-\Dtdivv_i,\,0\big\}.
\end{equation}
Here, $\xi_i$ is a limiter, detailed in \S\ref{sec:shear} below, aimed at
suppressing false detections of compressive flows in multi-dimensional flows.

\subsection{Adapting individual viscosities} \label{sec:AV:new}
Instead of increasing $\alpha_i$ by integrating a differential equation, we
set $\alpha_i$ directly to an appropriate local value
$\alpha_{\mathrm{loc},i}$ whenever this exceeds the current value for
$\alpha_i$. After extensive experimenting, we settled on the following
simple functional form
\begin{equation} \label{eq:a:loc}
  \alpha_{\mathrm{loc},i} =
  \alpha_{\max} \frac{h^2_iA_i}{\upsilon_{\mathrm{sig},i}^2+h^2_iA_i}
\end{equation}
with the signal velocity\footnote{Various definitions for the signal velocity
  can be found in the SPH literature. Ours reflects the maximum velocity with
  which information can be transported between particles, but avoids
  $\upsilon_{\mathrm{sig},i}\le0$.}
\begin{equation} \label{eq:vsig}
  \upsilon_{\mathrm{sig},i} = \max_{|\B{x}_{i\!j}|\le h_i}\big\{\bar{c}_{i\!j}
  -\min\{0,\B{\upsilon}_{i\!j}\cdot\hat{\B{x}}_{i\!j}\}\big\}.
\end{equation}
At the moment of passing through a shock (more precisely through a maximum of
the flow convergence), $A$ and hence $\alpha_{\mathrm{loc}}$ return to zero
and whenever $\alpha_i>\alpha_{\mathrm{loc},i}$ we let $\alpha_i$ decay
according to
\begin{equation}
  \label{eq:alpha:decay}
  \dot{\alpha}_i =  (\alpha_{\mathrm{loc},i}-\alpha_i)/\tau_i,
  \qquad
  \tau_i = h_i / 2\ell \upsilon_{\mathrm{sig},i}.
\end{equation}
We use $\upsilon_{\mathrm{sig},i}$ rather than $c$ in the decay time $\tau_i$
for internal consistency (this is of little practical relevance as
$\upsilon_{\mathrm{sig}}\approx c$ in the post-shock region). We use
$\ell=0.05$, such that the viscosity decays twice as slowly as in previous
methods, avoiding some occasional minor post-shock ringing not present in
methods with $\alpha_{\min}>0$. However, the traditional $\ell=0.1$ also gives
satisfactory results for most of our test problems.

\subsection{Avoiding false compression detections}
\label{sec:shear}
As explained in detail in Appendix~\ref{app:divv:fail}, in multi-dimensional
flows strong shear induces false detections of $\divv$ with the standard SPH
estimator even in the absence of particle disorder (noise). As shown in
Appendix~\ref{app:divv:good}, these errors can be reduced by first estimating
the velocity gradient matrix $\B{\s{V}}\equiv\B{\nabla}\B{\upsilon}$ and then
obtaining $\divv$ as its trace (we employ a similar method to estimate
$\Dtdivv$, see Appendices~\ref{app:divv:dot}).

Unfortunately, even with this improved method false detections for $\divv$
(and $\Dtdivv$) remain, for example in the situation of a differentially
rotating disc. These still induce artificial viscosity, which may be
significant in particular if $c_s/h$ is small compared to the shear. The
limiter $\xi_i$ in equation~(\ref{eq:A}) is aimed at suppressing such false
detections by $\xi_i\to0$ whenever the shear is much stronger than the
convergence \emph{and} no shock is present.

Having obtained the velocity gradient matrix $\B{\s{V}}\!$, the shear is easily
obtained as its traceless symmetric part
$\B{\s{S}}\equiv(\B{\s{V}}+\B{\s{V}}^t)/2 - \nu^{-1} (\divv) \B{\s{I}}$ (with
$\nu$ the number of spatial dimensions), while the presence of a shock is
indicated by 
\begin{equation} \label{eq:R}
  -1 \approx R_i \equiv \frac{1}{\hat{\rho}_i} \sum_j
  \mathrm{sign}(\divv_j)\,m_j\,W(|\B{x}_i-\B{x_j}|,h_i),
\end{equation}
since near a shock $\divv<0$ for all particles. After some experimenting, we
found the following functional form for the limiter suitable
\begin{equation} \label{eq:xi}
  \xi_i = \frac
      {|2(1-R_i)^4\,\divv_i|^2}
      {|2(1-R_i)^4\,\divv_i|^2 + \mathrm{tr}(\B{\s{S}}_i{\cdot}\B{\s{S}}_i^t)}.
\end{equation}
This functional form is similar to the Balsara limiter (\ref{eq:balsara}) in
that it compares the flow convergence to a measure of the traceless part of
the velocity gradient (the shear or the vorticity).

Alternatively, if one can be sure that no strong shear flows occur during the
simulation, one may use the standard SPH estimator for $\divv$ and estimate
$\Dtdivv$ from its change over the last time step. However, the limiter is
still desirable and one may use $|\curlv|^2$ instead of
$\mathrm{tr}(\B{\s{S}}{\cdot}\B{\s{S}}^t)$ in equation~(\ref{eq:xi}). We do
not use this simplified version in the tests presented below, but our
experiments indicate that such a method would pass all our tests except those
of \S\ref{sec:test:disc} and \S\ref{sec:test:shear}, both involving strong
shear.

\subsection{Behaviour in typical situations} \label{sec:AV:behaviour}
Before considering 2D and 3D test problems, we now assess the behaviour of our
novel scheme, as well as that of the M\&M method, in simple yet typical
situations.

First, consider a sound wave of velocity amplitude $\upsilon_s\ll c$ and wave
number $k\ll h^{-1}$ as example of a well-resolved weakly convergent flow. In
this case, $A\simeq k^2c\upsilon_s$ and $S\simeq k\upsilon_s$ at their
respective maxima. Since $\upsilon_{\mathrm{sig}}\simeq c\gg \upsilon_s$ we
have $\alpha_{\mathrm{loc}}\simeq\alpha_{\mathrm{max}}h^2k^2(\upsilon_s/c)$,
while for the M\&M method the asymptotic value
$\alpha_{\mathrm{s}}\simeq\alpha_{\min}+\alpha_{\max}hk(\upsilon_s/c)/2\ell$.
In the limit $kh\to0$ of a well-resolved wave, $\alpha_{\mathrm{loc}}\to0$
faster than $\alpha_{\mathrm{s}}\to\alpha_{\min}$, such that even with
$\alpha_{\min}=0$ the M\&M method would be more viscous than our new scheme.
Fig.~\ref{fig:damp_sound} shows 1D sound-wave SPH runs, demonstrating that our
new scheme behaves quasi-inviscid in this situation.

Following \cite{MorrisMonaghan1997}, we may also consider a simple homologous
flow $\B{\upsilon}=-a\B{x}$ with $a>0$, which approximates certain
astrophysical problems involving collapse and does not require artificial
dissipation. For this situation $S {\,=\,}3a$ but $A=0$ (a direct consequence
of the ability of $\Dtdivv$ to distinguish shocks from convergent flows),
such that our new scheme remains inviscid, while the M\&M method does not even
for $\alpha_{\min}=0$.

\begin{figure}
  \centerline{
    \resizebox{72mm}{!}{\includegraphics{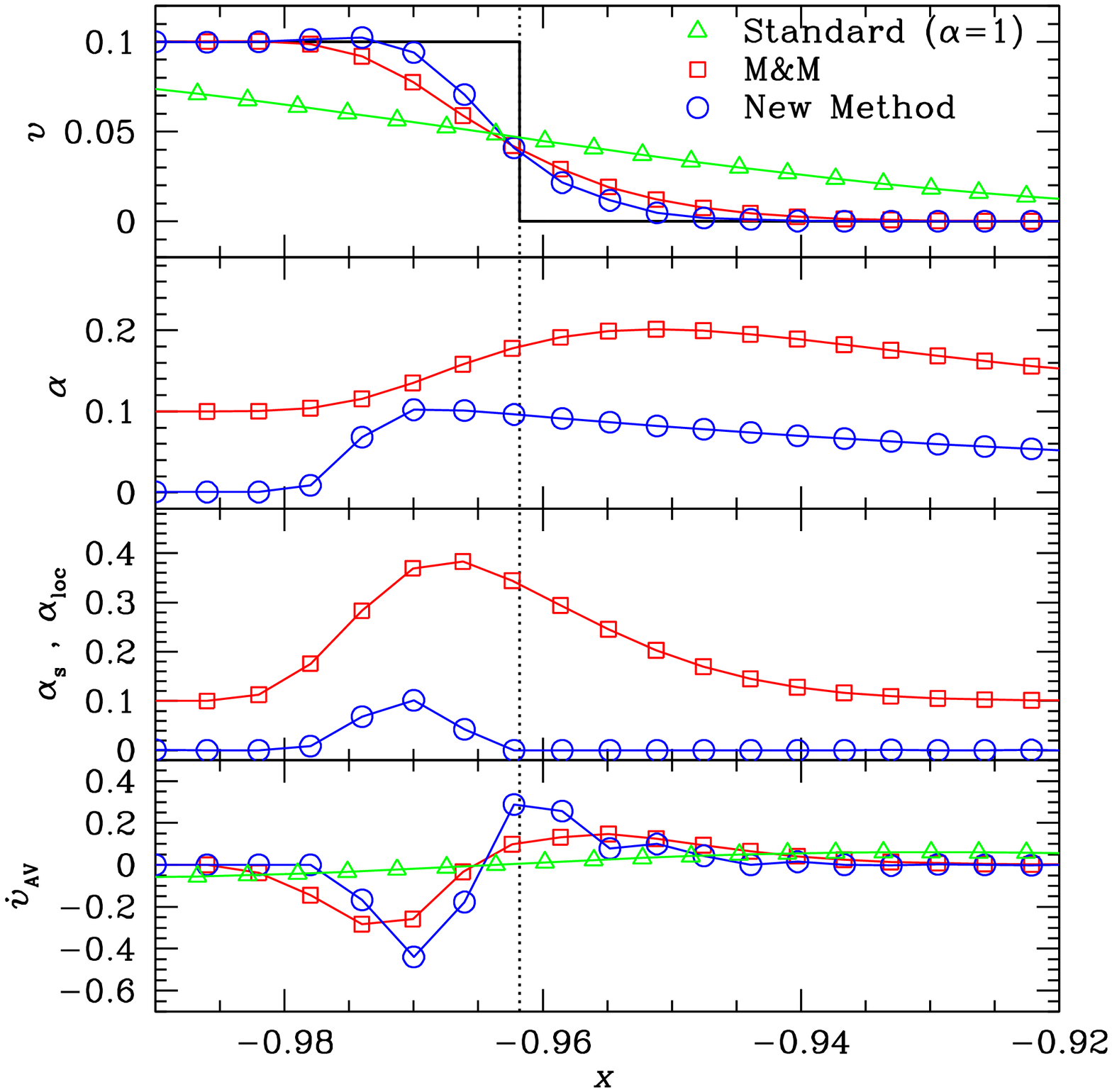}}
  }
  \centerline{
    \resizebox{72mm}{!}{\includegraphics{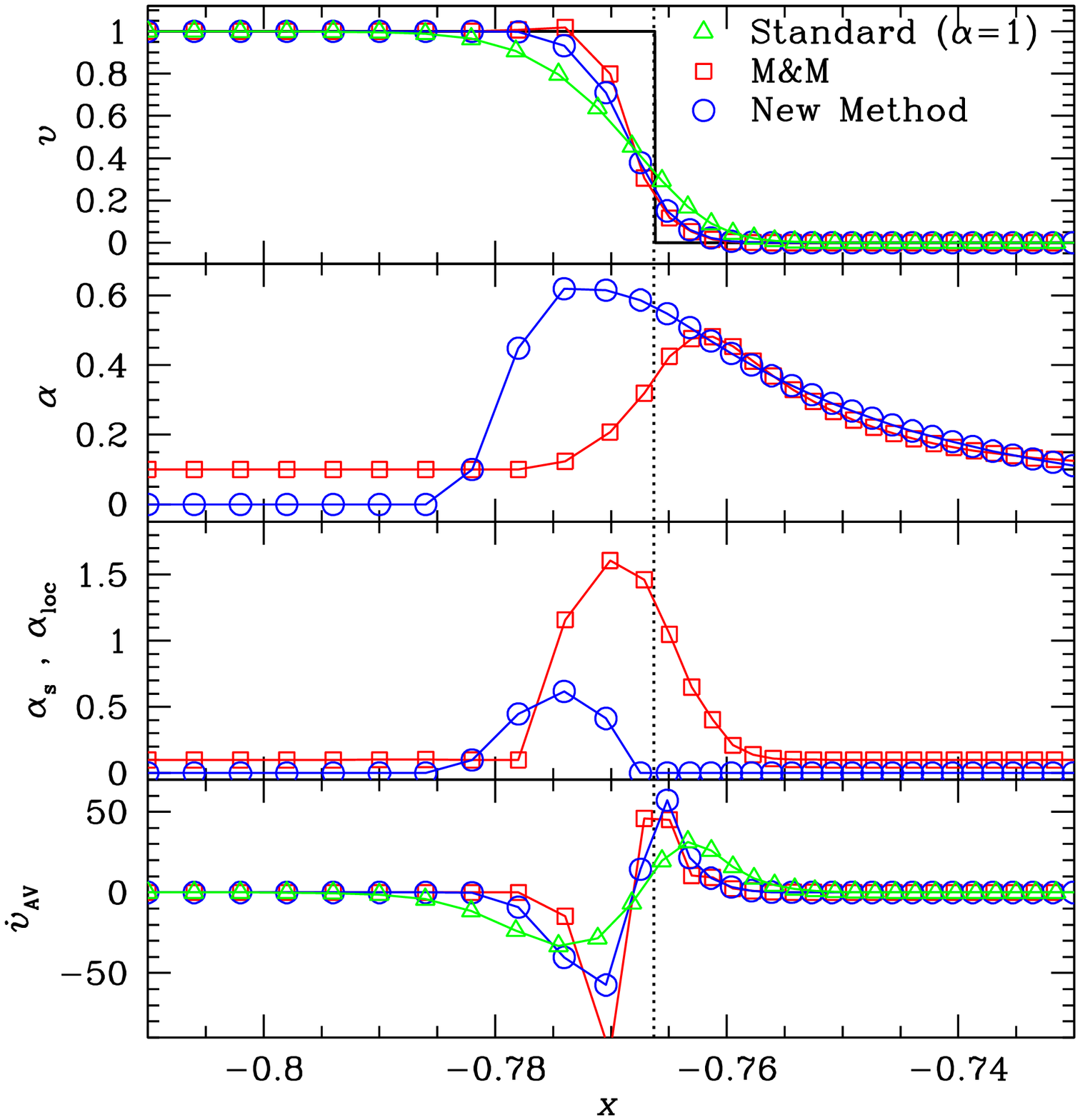}}
  }
  \caption{\label{fig:source} A 1D ram shock with $\delta\upsilon=0.1$
    (\textbf{top}) and $\delta\upsilon=1$ (\textbf{bottom}) in ideal gas
    with $\gamma=1.4$ simulated using standard SPH, the
    M\&M and our new method. We compare the velocity, viscosity parameter, its
    asymptotic value and the viscous deceleration. Initially, the velocity
    is discontinuous with $\upsilon=-\delta\upsilon\,\mathrm{sign}(x)$,
    resulting in two shocks of $\delta\upsilon$ propagating in either
    direction from the origin; the shock plotted propagates from right to
    left.}
\end{figure}

Next, consider a strong shock with velocity discontinuity $\delta\upsilon\gg
c$. Assuming that it is smoothed over one kernel width, we find maximum
amplitudes of $S\,{\simeq}\,\delta\upsilon/h$ and
$A\,{\simeq}\,(\delta\upsilon/h)^2$ (the exact values depend on the shock
conditions and the functional form of the smoothing kernel). Since
$\upsilon_{\mathrm{sig}}\simeq h \divv \sim \delta\upsilon$, our new scheme
gives $\alpha_{\mathrm{loc}} \sim \alpha_{\max}$, while the asymptotic value
(\ref{eq:a:asym}) for the M\&M method also approaches $\alpha_{\max}$.

While 3D simulations of strong shocks are presented in
\S\ref{sec:test:shock:strong}, Fig.~\ref{fig:source} presents weak ram-shock
simulations with $\delta\upsilon=0.1c$ (top) and $\delta\upsilon=c$ (bottom)
for our new scheme, the M\&M method, and standard SPH. In both regimes the
peak in, respectively, $\alpha_{\mathrm{loc}}$ and $\alpha_s$ is one particle
farther in front of the shock with our new method than with the M\&M method,
which reflects the superiority of $A$ over $S$ to detect an incoming
shock. This, combined with setting the viscosity parameter directly to the
required value, results in the peak in $\alpha$ to occur two particle
separations \emph{before} the shock for our new method, while for the M\&M
method it peaks a similar length \emph{behind} the shock.

With our new method, the viscous deceleration (bottom panels in
Fig.~\ref{fig:source}) sets in about three particle separations before the
weak and the strong shock, yielding good shock capturing properties in both
cases. The M\&M method, on the other hand, decelerates the flow much earlier
for a weak shock than for a strong shock and results in significant
over-damping of weak shocks (which also pertains to density and internal
energy -- not shown in Fig.~\ref{fig:source}), while our method smoothes both
shocks over four particle separations (top panels in Fig.~\ref{fig:source}),
the optimal SPH resolution in 1D. Note that standard SPH is hopeless: it
over-smoothes the strong shock and is completely incapable of dealing with the
weak shock.

\subsection{Maintaining particle order}
\label{sec:AV:amin}
The main point of our method is the absence of artificial viscosity away
from shocks. Hence, if $\alpha_{\min}>0$ was indeed required to maintain
particle order, as previously argued in context of the M\&M method, our
method should fail in this regard. Noise in SPH can emerge from shocks or
carelessly generated initial conditions.

\begin{figure}
  \centerline{
    \resizebox{80mm}{!}{\includegraphics{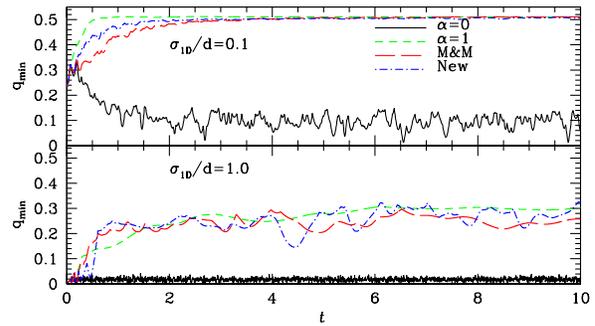}}
  }
  \caption{\label{fig:noise} Time evolution of $q_{\min}$, defined in
    equation~(\ref{eq:qmin}), for SPH simulations started from noisy initial
    conditions (see text). All SPH schemes with artificial viscosity suppress
    the noise equally well.}
\end{figure}
Let us first consider the time evolution of noisy initial conditions,
generated by adding random displacements to particle positions representing
noise-free hydrostatic equilibrium (the vertices of a face-centred-cubic grid,
i.e.\ densest-sphere packing). We consider two cases with the displacements in
each dimension drawn from a normal distribution with rms amplitude equal to
the nearest-neighbour distance or a tenth of it, respectively. The time
evolution of such noisy initial conditions can be distinguished by monitoring
\begin{equation} \label{eq:qmin}
  q_{\min} \equiv \min_{i,j}\big\{|\B{x}_{i\!j}|/h_i\big\}.
\end{equation}
There are three possible scenarios. Either the particles settle back close to
the original grid ($q_{\min}$ approaches its grid value $q_{\mathrm{grid}}$),
form a glass ($q_{\min}$ approaches a finite value $<q_{\mathrm{grid}}$), or
form dense clumps (`clumping instability',
$q_{\min}\sim0$). Fig.~\ref{fig:noise} plots the evolution of $q_{\min}$ for
$N_h=40$ SPH neighbours (see also Appendix~\ref{sec:adapt}) when
$q_{\mathrm{grid}}\approx0.529$. Clumping only occurs when $\alpha\equiv0$,
while for any viscous scheme tested the particles settle back onto the grid or
form a glass with roughly similar time evolutions.

\begin{figure}
  \centerline{
    \resizebox{80mm}{!}{\includegraphics{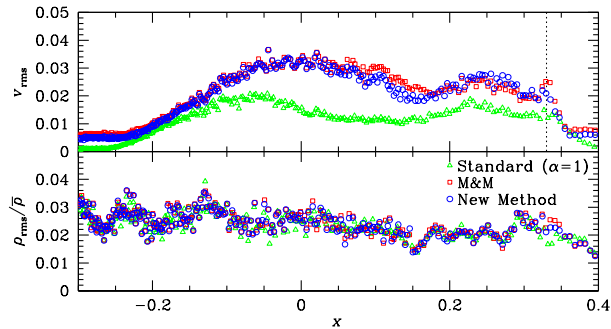}}
  }
  \caption{\label{fig:sod:noise} The rms amplitudes of density and velocity
    fluctuations for 3D simulations of the
    Sod (1978)
    shock tube test (see also Fig.~\ref{fig:sod}). Initial conditions were
    prepared using a glass. The shock propagates to the right and is indicated
    by the dotted line; the velocity jump at the shock is 0.63.}
\end{figure}
Post-shock noise occurs because the shock-induced compression disrupts the
original particle order, but other than in the above test the viscosity is
already switched on. In Fig.~\ref{fig:sod:noise}, we plot the amplitudes of
the velocity and density noise in 3D simulations of the standard
\cite{Sod1978} shock tube test (see also \S\ref{sec:sod}). The three methods
have similar levels of density noise, but standard SPH is less noisy in the
velocities, which is not surprising given its stronger viscosity. However,
between the two viscosity suppressing schemes there is little difference, even
though $\alpha_{\min}=0$ for our method. Similar results obtain for other
shock tests and we conclude that our method is no worse than M\&M's for
maintaining particle order.

\section{Viscosity Suppression Tests}
\label{sec:test:suppress}
We now present some tests of low-Mach-number flows, where previous methods
give too much unwanted dissipation.

\begin{figure}
  \centerline{
    \resizebox{75mm}{!}{\includegraphics{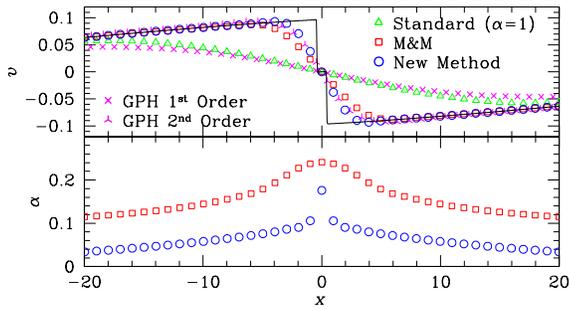}}
  }
  \caption{\label{fig:1D_sound} Steepening of a 1D sound wave: velocity and
    viscosity parameter vs.\ position for standard SPH, the M\&M method, our
    new scheme, and Godunov particle hydrodynamics of first and second order
    \citep[GPH,][]{ChaWhitworth2003}, each using 100 particles per
    wavelength. The solid curve in the top panel is the solution obtained with
    a high-resolution grid code.}
\end{figure}

\subsection{Sound-wave steepening}
\label{sec:test:soundwave}
The steepening of sound waves is a simple example demonstrating the importance
of distinguishing between converging flows and shocks. As the wave propagates,
adiabatic density and pressure oscillations result in variations of the sound
speed, such that the density peak of the wave travels faster than the trough,
eventually trying to overtake it and forming a shock.

In our test, a 1D sound wave with a velocity amplitude $10\%$ of the sound
speed is used (ideal gas with $\gamma=1.4$). Fig.~\ref{fig:1D_sound} compares
the velocity field at the moment of wave steepening for various SPH schemes,
each using 100 particles, with a high-resolution grid simulation. The new
method resolves the shock better than the M\&M scheme, let alone standard SPH.

In Fig.~\ref{fig:1D_sound}, we also show results from GPH \citep[Godunov-type
  particle hydrodynamics,][]{ChaWhitworth2003}, which differs from SPH by
using the pressure $P^*$, found by solving the Riemann problem between
particle neighbours, in the momentum and energy equations and avoids the need
for explicit artificial viscosity. This substitution does not affect the
energy or momentum conservation \citep{Cha2002}, and indeed we find that both
are well conserved. While the first-order GPH scheme is comparable to standard
SPH and also to an Eulerian Godunov grid code using the same Riemann solver
without interpolation (not shown), the second-order GPH scheme resolves the
discontinuity almost as well as our novel method.

\begin{figure}
  \centerline{
    \resizebox{74mm}{!}{\includegraphics{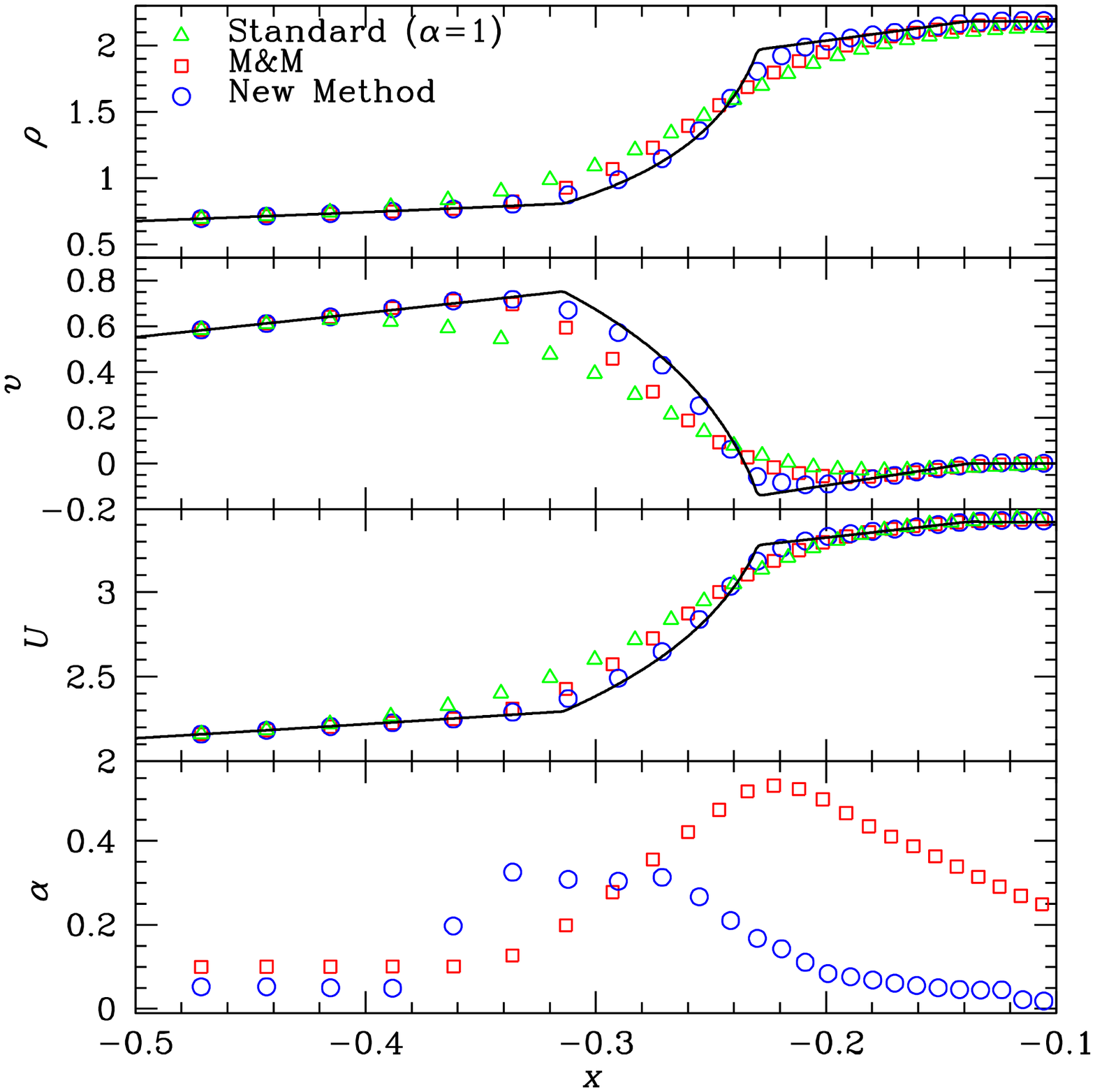}}
  }
  \centerline{
    \resizebox{74mm}{!}{\includegraphics{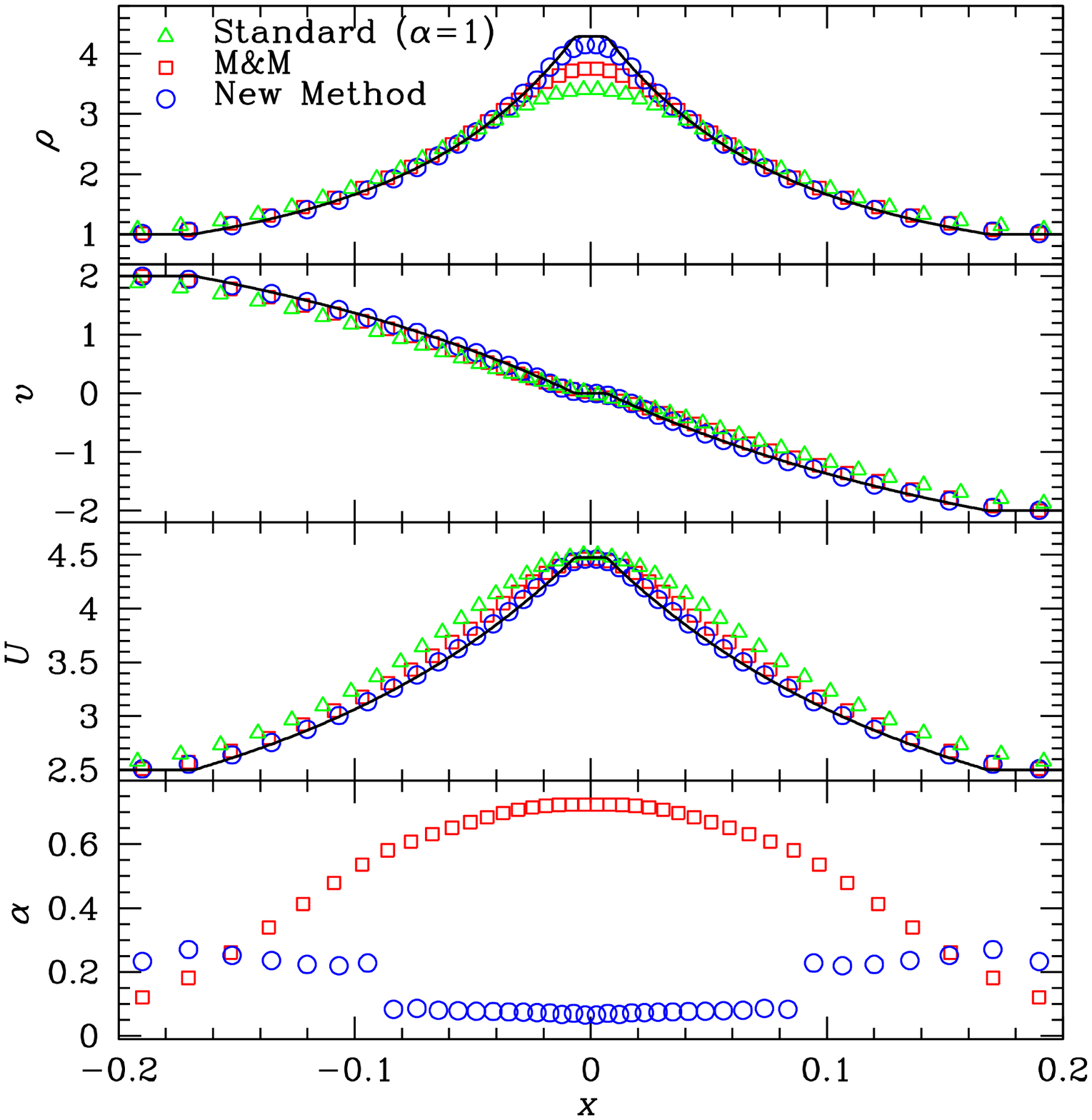}}
  }
  \caption{\label{fig:converging} A 1D converging flow test with initially
    constant density and pressure and velocities given by equation
    (\ref{eq:converging_ini}) using an adiabatic equation of state with
    $\gamma=1.4$. \textbf{Top}: run for $\upsilon_a=1$ at $t=0.3$;
    \textbf{bottom}: run for $\upsilon_a=2$ at $t=0.1$. The solid lines are the
    result of a high-resolution Eulerian grid-code simulation.}
\end{figure}
\begin{figure*}
  \centerline{
    \resizebox{150mm}{!}{\includegraphics{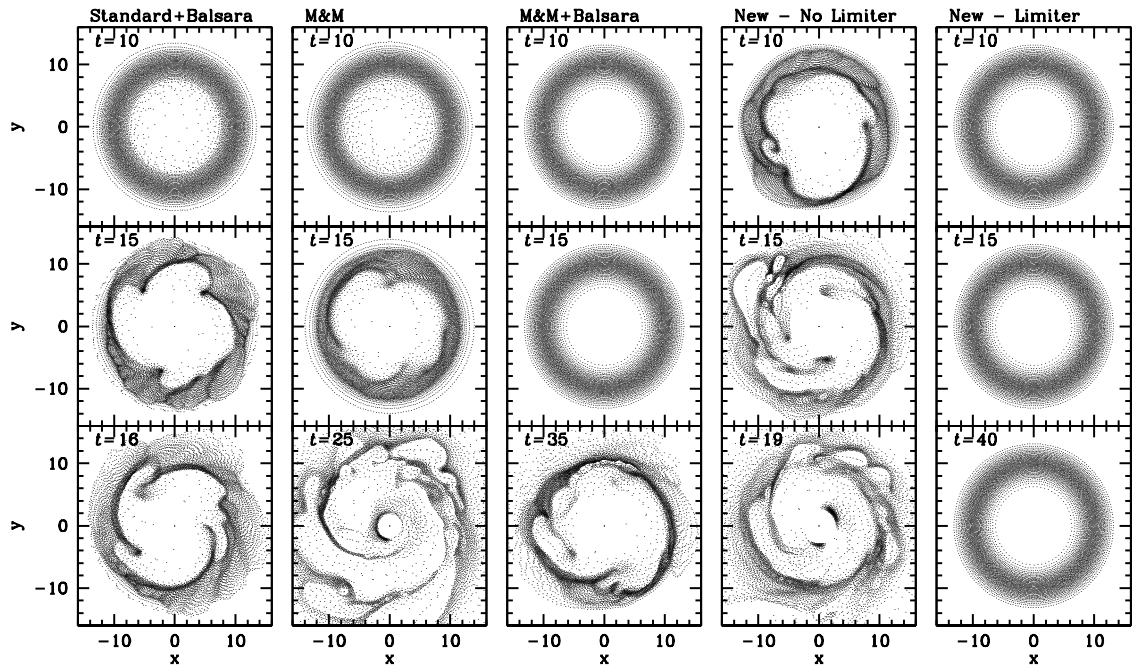}}
  }    
  \caption{\label{fig:ring} Keplerian ring test: particle positions at various
    times for standard SPH with Balsara switch, the M\&M method with and
    without Balsara switch, and our new method without and with the viscosity
    limiter $\xi$ of equation~(\ref{eq:xi}). Only for this last method the
    ring remains stable against a viscosity-induced instability. (Ring-like
    features at $r\lesssim2$ are artifacts caused by the dynamical time close
    to the centre being short compared to the time step).  }
\end{figure*}

\subsection{1D converging flow test}
Similar to sound-wave steepening, this test requires good treatment of
convergent flows and weak shocks. The initial conditions are uniform pressure
and density and a continuous flow velocity
\begin{equation}
  \label{eq:converging_ini}
  \upsilon = \begin{cases}
    4(1+x)\upsilon_a &-1.00<x<-0.75, \\
    \upsilon_a       &-0.75<x<-0.25, \\
    -4x\upsilon_a    &-0.25<x<\phantom{-}0.25, \\
    -\upsilon_a      &\phantom{-}0.25<x<\phantom{-}0.75, \\
    4(1-x)\upsilon_a &\phantom{-}0.75<x<\phantom{-}1.00.
  \end{cases}
\end{equation}
As there is no analytical solution, we compare the results to a
high-resolution grid-code simulation. We run tests for $\upsilon_a=1$ and
$\upsilon_a=2$ as shown in the top and bottom panels of
Fig.~\ref{fig:converging}.

While the M\&M switch certainly improves upon standard SPH, it still
over-smoothes the velocity profile as the viscosity is increased before a shock
has formed. This is particularly evident in the velocity profile of the
$\upsilon_a=2$ case (bottom) near $x=0$. The new switch keeps the viscosity low,
in the $\upsilon_a=2$ case an order of magnitude lower than the M\&M method. In
fact, the agreement between our method and the high-resolution grid code is as
good as one can possibly expect at the given resolution, in particular the
velocity plateau and density amplitude around $x=0$ in the $\upsilon_a=2$ case
(bottom) are correctly modelled.

\subsection{2D Keplerian-ring test} \label{sec:test:disc}
In this test, a gaseous ring orbits a central point mass, neglecting the
self-gravity of the gas. Initially, the ring is in equilibrium: pressure
forces, attraction by the point mass, and centrifugal forces balance each
other. The Keplerian differential rotation implies that the flow is shearing
and any viscosity causes the ring to spread
\citep{Lynden-BellPringle1974}. This is indeed what
\cite*{MaddisonMurrayMonaghan1996} found in SPH simulations without pressure
forces.

\citeauthor{MaddisonMurrayMonaghan1996} also found an instability to develop
from the inner edge, which quickly breaks up the ring. They argue convincingly
that this is the viscous instability \citep*{LyubarskijPostnovProkhorov1994},
which causes eccentric orbits at the inner edge of the ring to become more
eccentric due to the viscous deceleration peaking at apo-centre.

\cite{ImaedaInutsuka2002} performed SPH simulations of the same problem but
including pressure forces. They find a similar break-up of the ring after only
few rotations and blame it on an inadequacy of the SPH scheme itself. We
strongly suspect that \citeauthor{ImaedaInutsuka2002} encountered a form of
the clumping instability, which appears to be particularly strong in 2D
simulations of strong shear flows \citep[though it may have been a dynamical
  instability inherent to gaseous Keplerian rings,
  e.g.][]{PapaloizouPringle1984,PapaloizouPringle1985,GoldreichNarayan1985}.
This numerical instability grows on a local hydrodynamical time and may
therefore be suppressed by choosing the sound speed $c$ much lower than the
rotation speed $\upsilon_\varphi$. Indeed, \cite{Price2004} and
\cite{Monaghan2006}, who repeated these and similar experiments with a very
low sound speed, found no such numerical instabilities. A detailed
investigation of these issues is clearly beyond the scope of our study and we
merely compare our new scheme to previous methods for pressure forces with
$c\ll\upsilon_\varphi$ when the viscous instability should strike after few
rotations depending on the strength of the artificial viscosity.

In our test, $GM=1000$ for the central point mass, while the gas ring has
Gaussian surface density centred on $r=10$ with width (standard deviation)
$2.5$ represented by $N=9745$ particles initially placed according to the
method of \cite*{CartwrightStamatellosWhitworth2009}. This implies an orbital
period of $2\pi$ and velocity of $\upsilon_\varphi=10$ at the ring centre. We
choose a sound speed of $c=0.01\ll\upsilon_\varphi$ to ensure that any
dynamical instabilities of inviscid rings become important only after many
periods.

Figure \ref{fig:ring} shows the particle distributions at various times for
different SPH schemes. Only with our new method, the rings stay in their
initial equilibrium configuration over at least five periods, while for the
other methods, the inner parts of the ring soon become disordered leading to a
catastrophic break-up after a few periods. It is noteworthy that this failure
occurs despite the Balsara switch, which was designed specifically for
applications like this.

Note that without the viscosity limiter $\xi$ of equation~(\ref{eq:xi}), our
novel method fails, precisely because of shear causing false detections of
flow compression (as highlighted in \S\ref{sec:shear} and
Appendix~\ref{app:divv}).

\begin{figure*}
  \centerline{
    \resizebox{68mm}{!}{\includegraphics{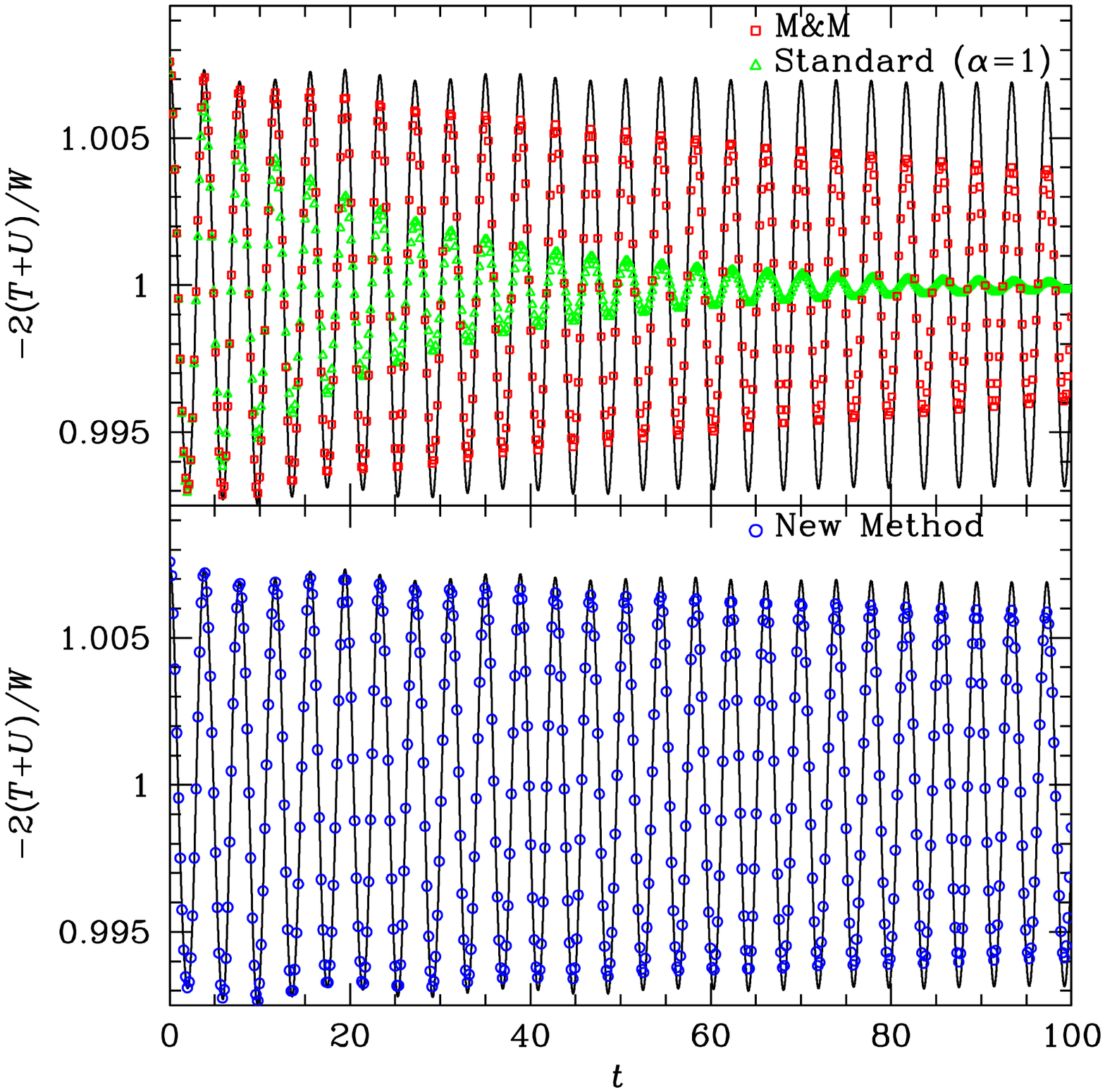}}
    \hfil
    \resizebox{68mm}{!}{\includegraphics{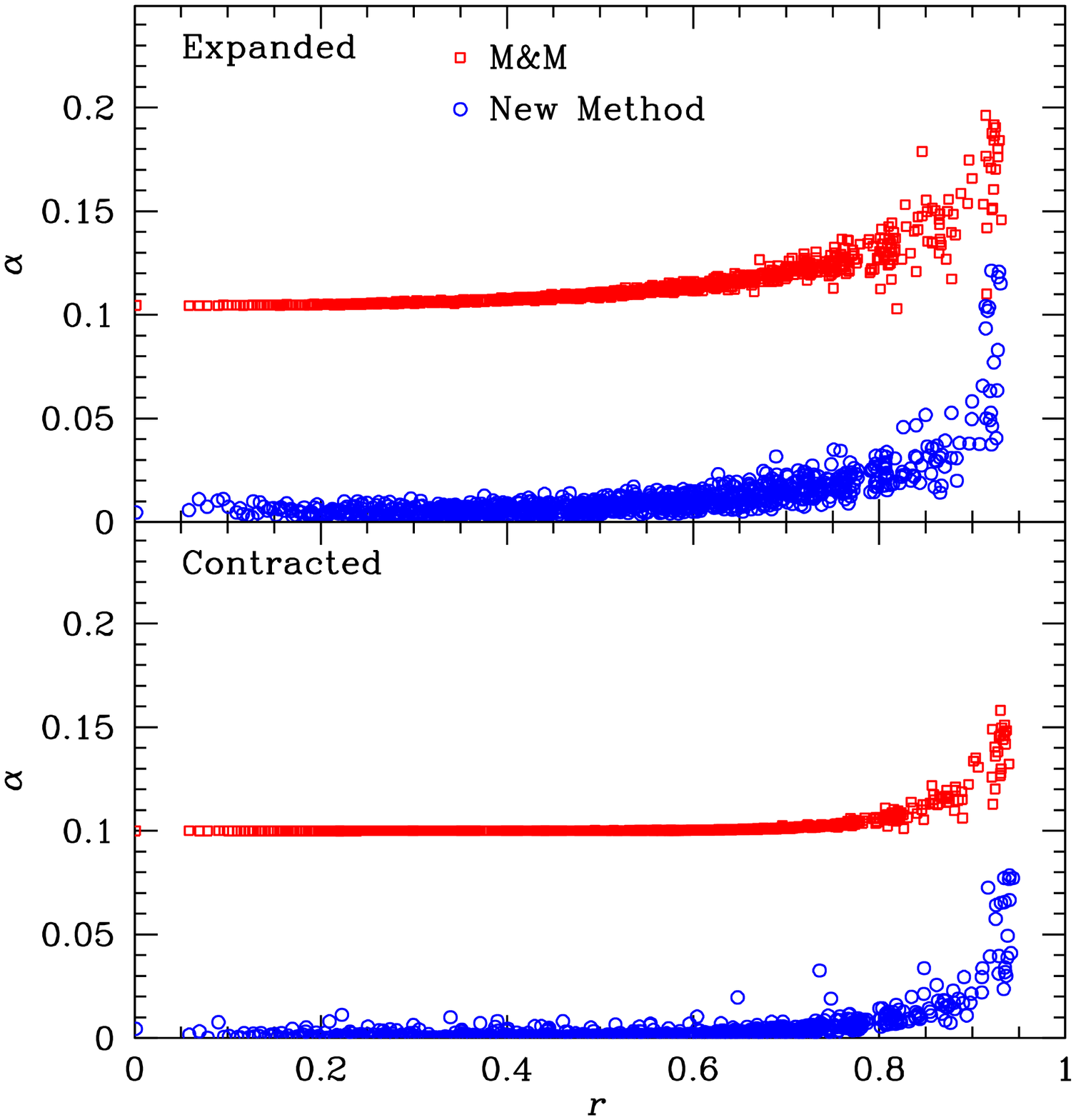}}
  }
  \caption{\label{fig:polyosc} \textbf{Left}: virial ratio plotted versus time
    for SPH models of a radially oscillating polytrope which initially was in
    its fundamental radial eigenmode with relative radial amplitude of $0.01$
    and period $3.89$. The \emph{solid} curves are for a SPH model without any
    artificial viscosity. \textbf{Right}: the viscosity parameter $\alpha$ at
    $t=97$ (maximum contraction) and $t=99$ (maximum expansion) for every
    100th particle. The new method keeps viscosity lower at the edge of the
    polytrope.}
\end{figure*}
\begin{figure*}
  \centerline{
    \resizebox{68mm}{!}{\includegraphics{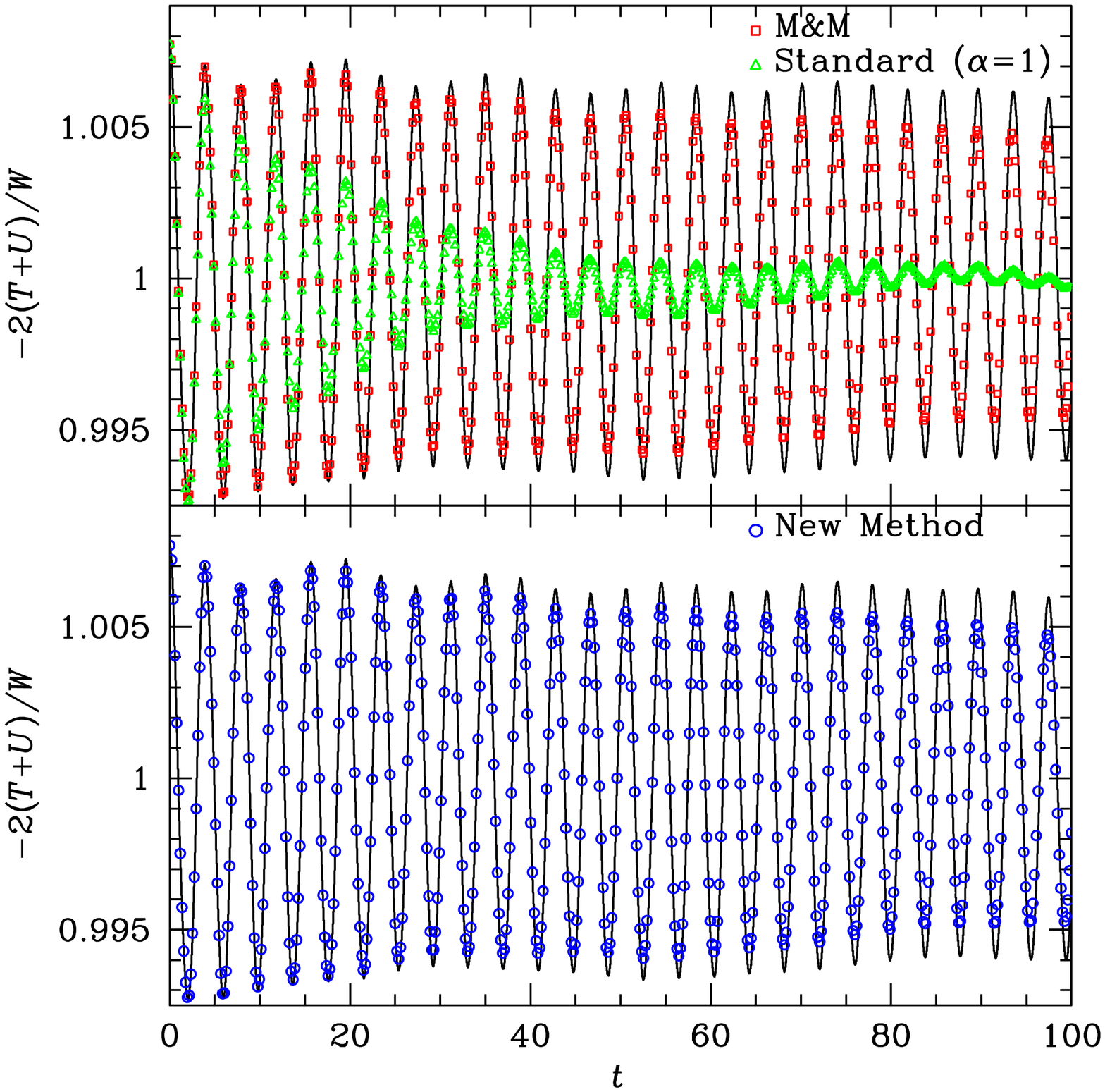}}
    \hfil
    \resizebox{68mm}{!}{\includegraphics{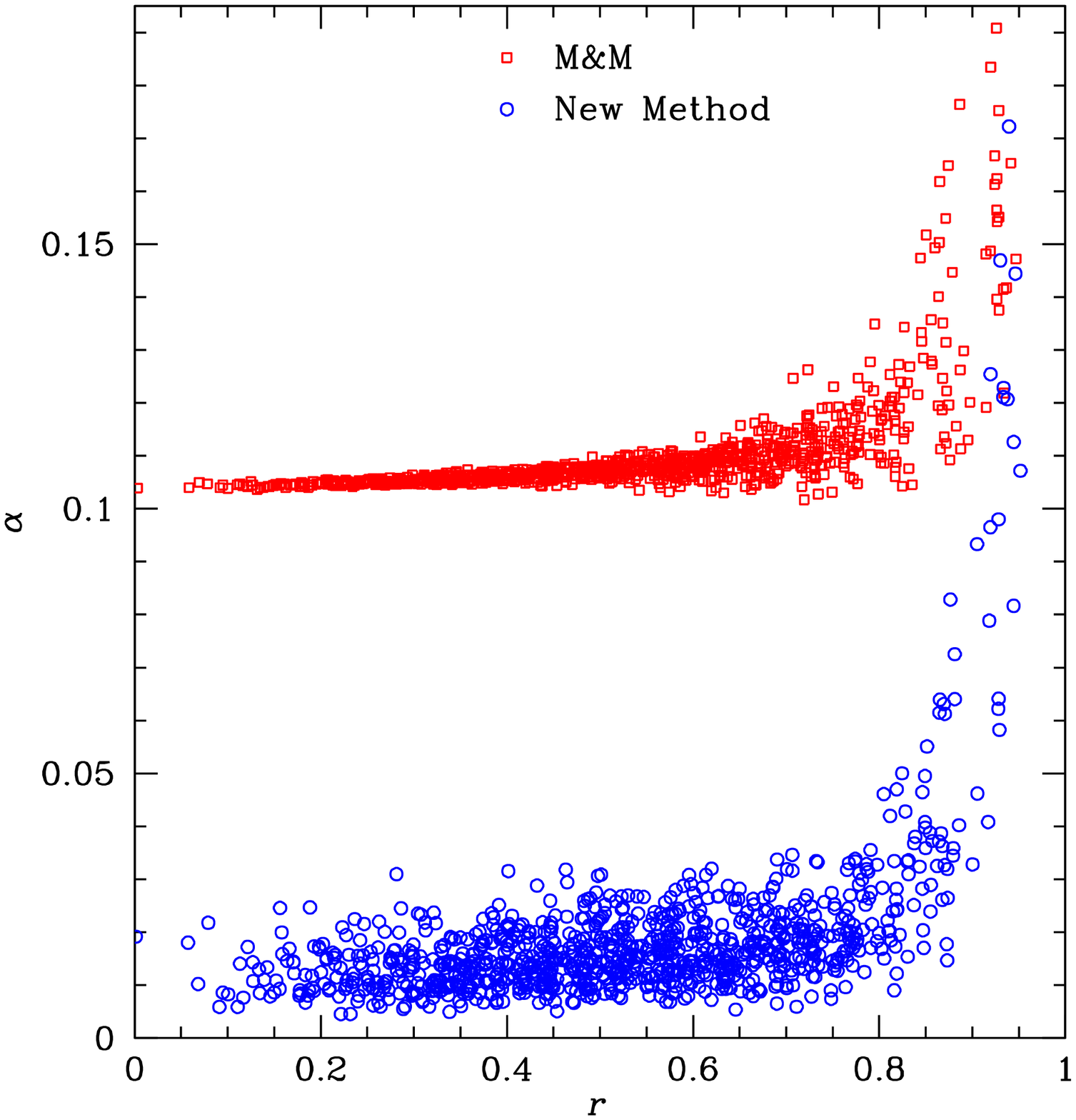}}
  }
  \caption{\label{fig:polypoint} Same as Fig.~\ref{fig:polyosc}, except that
    the sphere is in circular orbit around a point mass of 100 times its mass
    and with orbital radius 20 times its radius (the kinetic and potential
    energies are corrected for the contributions from the orbit). The
    viscosity parameter for every 100th particle is plotted at $t=100$
    (\textbf{right}).}
\end{figure*}
\begin{figure*}
  \centerline{
    \resizebox{140mm}{!}{\includegraphics{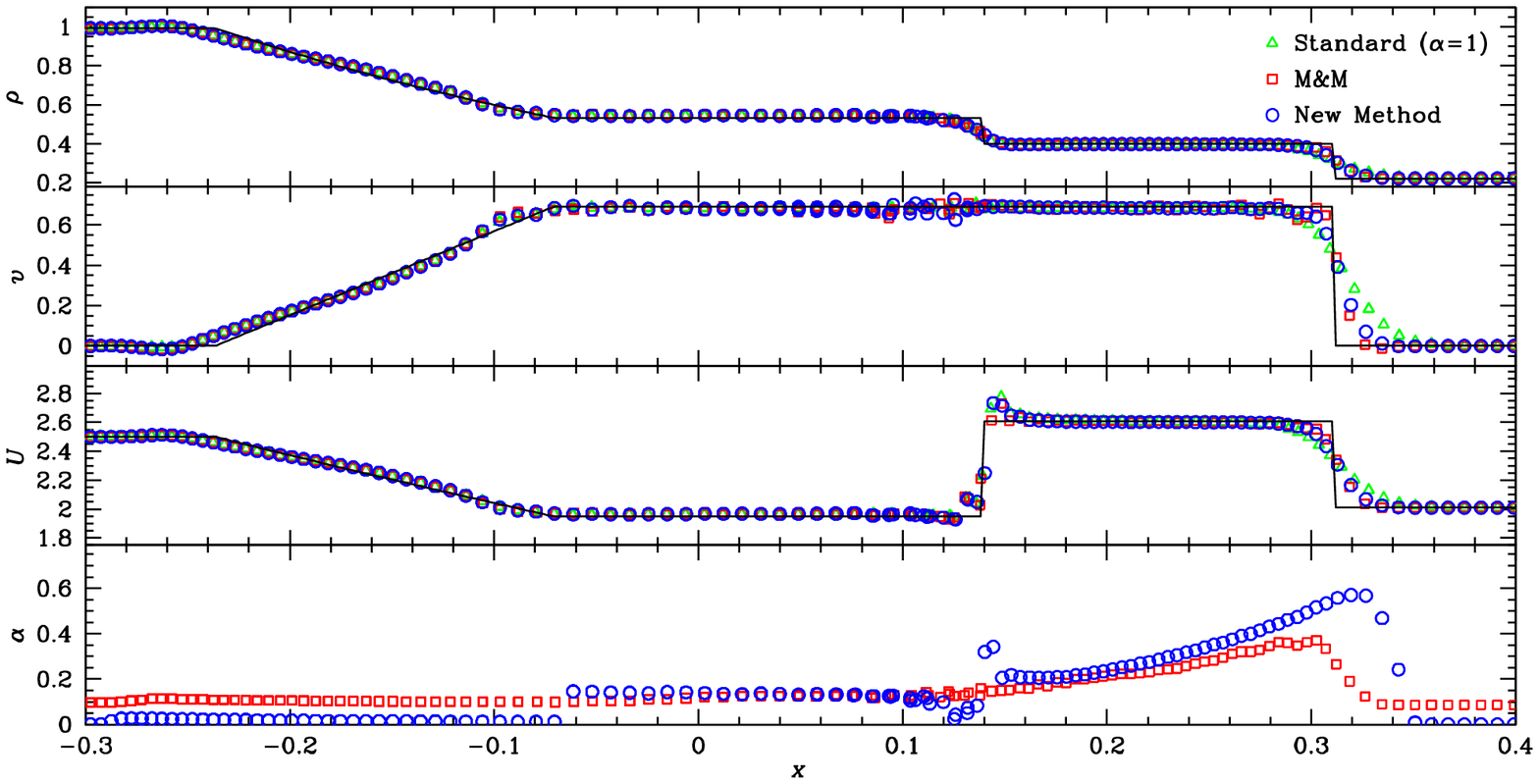}}
  }
  \caption{\label{fig:sod} Comparison of our new scheme and the M\&M method
    for the standard
    Sod (1978)
    shock tube test with the analytic solution (\emph{solid}).}
\end{figure*}

We also run similar tests with the central point mass replaced by a mass
distribution (Plummer sphere or Kuzmin disc) with gravitational potential
$\Phi=-GM/\sqrt{r^2+s^2}$ with $s=3$, such that the rotation curve of the disc
also contains a rising part, similar to the situation in galactic discs. The
outcome of these simulations (not shown) is essentially identical to that for
the pure Keplerian rings: only our new method with viscosity limiter does not
fall prey to the viscous instability.

\subsection{An oscillating polytropic sphere}
\label{sec:test:poly}
The pulsations of a polytropic sphere are a good test for the adverse effects
of artificial dissipation \citep{SteinmetzMueller1993}. We set up a polytropic
sphere of $10^5$ particles and induce oscillations in its fundamental mode
\citep[e.g.][]{Cox1980} with relative amplitude of $0.01$ in radius and a
period of $P=3.8$.

In the absence of viscosity we expect the radial oscillations to continue with
the initial amplitude and period over many oscillations. However, as with any
numerical method some small amount of numerical dissipation may appear.
Nonetheless, such effects should be small compared to the dissipation caused
by artificial viscosity. Since the size of the radial perturbations increases
with radius, we expect the oscillations to be small at the centre of the
polytrope and therefore our new method to keep the viscosity low there.
However, at the edge the size of the oscillations is more significant, and we
may see an increase in viscosity at this point.

In order to track the oscillations, we monitor in Fig.~\ref{fig:polyosc} the
time evolution of the virial ratio $-2(T+U)/W$ where $T$, $U$, and $W$, are
the kinetic, internal, and the gravitational energies, respectively. At
maximum contraction the virial ratio is at its peak and at maximum expansion
the virial ratio is lowest. With no artificial viscosity (solid curves in
Fig.~\ref{fig:polyosc}) the wave remains at constant amplitude barring
a slight initial drop. The period averaged over 25
oscillations is $P=3.89$, only slightly larger than the expected value. The
reason for this discrepancy is most likely the unavoidable deviation of the
(finite-resolution) SPH model from a perfect polytropic sphere. This deviation
also means that our SPH model is not exhibiting a pure eigenmode, but in
addition contains some higher-order modes at low amplitudes, resulting in some
beating between them.

The M\&M method results in a slow but continuous decay of the oscillations,
though the period is hardly affected. This damping can be blamed largely on
the finite $\alpha_{\min}$ (standard SPH damps the oscillation ten times
faster). Conversely, our new method, hardly damps the oscillations at
all, because $\alpha$ is kept very small (except for the outermost layers
where $\alpha$ is still below the M\&M values).

We also run simulations where the oscillating polytropic sphere is on a
circular orbit 20 times the radius of the sphere around a point mass 100 times
that of the sphere (corresponding to a period of 56 time units). With this
choice, the tidal radius is approximately four times the radius of the gas
sphere, implying that tides are strong but not catastrophic. Since the orbital
accelerations are much larger than those due to the polytropic oscillations,
this is a tough test for any numerical scheme. In particular, Eulerian methods
should have severe problems (this does exclude using co-rotating coordinates,
which do not allow for tidal evolution of the orbit and are unavailable for
eccentric orbits).

The time evolution of the virial ratio and the viscosity parameter $\alpha$
are shown in Fig.~\ref{fig:polypoint} for the same viscosity schemes as for
the isolated case in Fig.~\ref{fig:polyosc}. First note that the undamped
simulations (solid curves) behave differently from the isolated case,
exhibiting variations and a slight decay, both of which are most likely caused
by the tidal field. As to be expected for any Lagrangian scheme, both SPH
methods perform very similar to the isolated case, because neither $\divv$ nor
$\Dtdivv$ are affected by the orbital acceleration.

\section{Shock capturing tests}
\label{sec:test:shock}
In this section, we subject our method to situations where artificial
viscosity is required, mainly high-Mach number shocks, and our aim is to
demonstrate that it performs at least as well as the M\&M method.

\subsection{Sod shock tube test} \label{sec:sod}
The \cite{Sod1978} shock tube test is a standard test for any shock capturing
method and consists of an initial discontinuity in pressure and density
leading to the production of a rarefaction wave, contact discontinuity and
shock wave, which forms from the steepening of a subsonic wave. The whole
system is subsonic with a maximum Mach number of $\mathcal{M} \approx 0.63$ in
the pre-shock region. We perform the test in 3D at a resolution of 200
particle layers in the high-density region.

The density, energy, velocity, and viscosity for standard SPH as well as the
M\&M and our method are shown in Fig.~\ref{fig:sod}. As for the 1D ram test
(see Fig~\ref{fig:source}), our new method switches on viscosity already in
the pre-shock region peaking about one smoothing length before the actual
shock front (which travels to the right in Fig.~\ref{fig:sod}), whereas the
M\&M switch turns on viscosity later, lagging our method by about four
particle separations. As a consequence, the transition of the fluid values
across the shock front is slightly smoother with our method than with the M\&M
method.

Note that the irregularities around the contact discontinuity at $x=0.138$
common to all schemes tested are not related to artificial viscosity (the
irregularities in $\alpha$ at that point could be removed by choosing non-zero
initial $\alpha$ at the initial discontinuity); they can be alleviated by
artificial conductivity \citep{Price2004,Price2008}.

\begin{figure}
  \centerline{
    \resizebox{70mm}{!}{\includegraphics{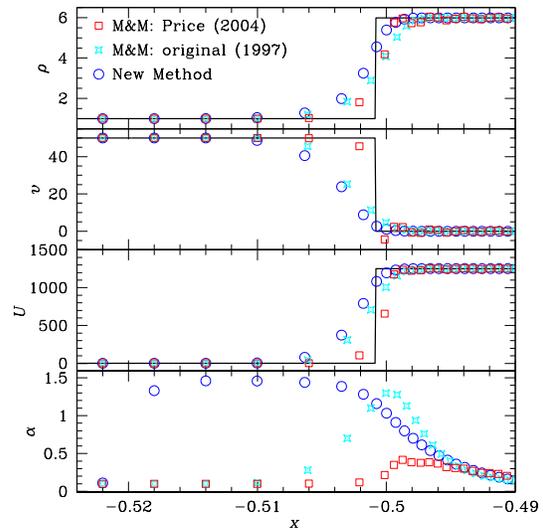}}
  }
  \caption{\label{fig:1Dram50} Same as Fig.~\ref{fig:source} but for
    $\mathcal{M}=50$. We distinguish between the original M\&M method (using
    eq.~\ref{eq:davdt}) and the
    Price (2004)
    version (using eq.~\ref{eq:davdt:alt} with $\alpha_{\max}=2$), which has
    been denoted `M\&M' in all figures so far.}
\end{figure}

\subsection{Strong shocks and particle penetration}
\label{sec:test:shock:strong}
In \S\ref{sec:AV:behaviour} and Fig.~\ref{fig:source} we already demonstrated
that our new method is superior to the M\&M scheme in resolving `subsonic
shocks' (velocity discontinuities smaller than the sound speed) and comparable
in resolving shocks of Mach number $\sim1$. Here, we extend this comparison to
high Mach numbers. Fig.~\ref{fig:1Dram50} shows the result for the 1D ram test
with $\mathcal{M}=50$. The \cite{Price2004} version of the M\&M method, which
uses equation~(\ref{eq:davdt:alt}) with $\alpha_{\max}=2$, is implemented in
some contemporary SPH codes, and has been used in our tests so far, fails this
test: $\alpha$ remains too low and as a consequence the velocity discontinuity
is not correctly smoothed and some post-shock ringing occurs. To give credit
to \cite{MorrisMonaghan1997}, we also tested their original method and find it
to work well (stars in Fig.~\ref{fig:1Dram50}).  Our new method works about as
well as the original M\&M scheme, with $\alpha$ reaching the same level,
though our scheme detects the coming shock much earlier: $\alpha$ is ahead of
the original M\&M method by about four particle separations.

Whilst the main role of artificial viscosity is to resolve shocks by
transferring entropy, a secondary but vital role is to prevent particle
penetration, which requires strong viscosity in high Mach number shocks.
\cite{Bate1995} performed many tests to determine the value of the parameters
$\alpha$ and $\beta$ needed to prevent particle penetration in ram shock tests
of various Mach numbers. For particles arranged in face-centred-cubic or
cubic grids, Bate found that appropriate values for the viscosity parameters
can prevent particle penetration for shocks up to $\mathcal{M}=8$.  Most SPH
practitioners opt for a value of $\beta=2\alpha$ \citep{MorrisMonaghan1997}.

\begin{figure}
  \centerline{
    \resizebox{70mm}{!}{\includegraphics{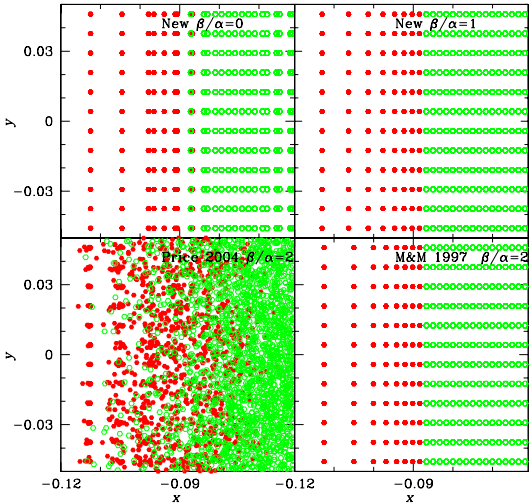}}
  }
\caption{\label{fig:3Dram} Particle positions in the $x$-$y$ plane of 3D
  simulations of a $\mathcal{M}=20$ ram shock along $x$ direction. Particles
  are coloured red if there initial positions was $x_0<-0.45$ and green if
  $x_0>-0.45$}
\end{figure}

To determine the correct value of $\beta$ required for the new scheme, we
perform high-resolution 3D runs of ram shocks with $\mathcal{M}{\,=\,}20$ and
various values for $\beta/\alpha$. We smooth the initial velocity
discontinuity, as suggested by \cite{Monaghan1997}, to provide the method with
a situation realistic for SPH, such as would have arisen for a shock forming
from continuous initial conditions.

For different values of $\beta/\alpha$ with our viscosity scheme and the two
variants of the M\&M switch, we plot in Fig.~\ref{fig:3Dram} the $x$ and $y$
positions (for all values of $z$) of particles near the shock front at a late
time. The colour coding distinguishes particles which at that time should be
up- (red) or downstream (green). Our scheme prevents particle penetration with
$\beta=\alpha$ (for $\beta=0$ there is one layer of overlap). The original
M\&M scheme with the standard choice $\beta=2\alpha$ also avoids particle
penetration, but not the \cite{Price2004} version, again a consequence of too
little viscosity.

\subsection{A shearing shock}
\label{sec:test:shear}
\begin{figure}
  \centerline{ \resizebox{70mm}{!}{\includegraphics{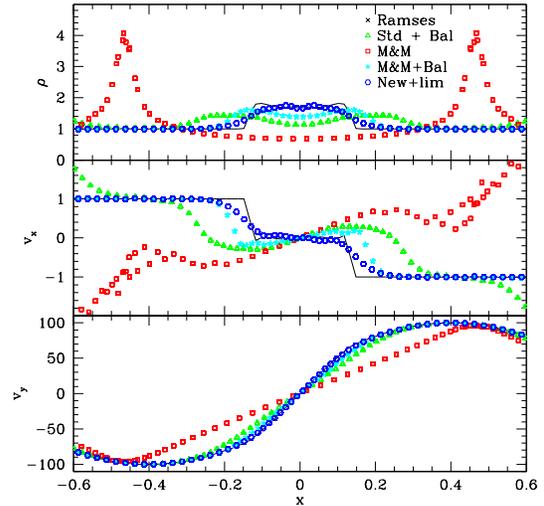}}}
  \caption{\label{fig:shear:shock} Shearing shock test: density and
    velocity for various SPH schemes (symbols) and a grid-code simulation
    (curve). Initial velocities are given by
    equation~(\ref{eq:shear:shock}) with $s=100\upsilon=100c$.}
\end{figure}

This test combines a shock with a perpendicular shear and presents a difficult
test for any SPH scheme. We use periodic boundary conditions and start from
a face-centred cubic grid and velocities
\begin{equation} \label{eq:shear:shock}
  \upsilon_x = -\delta\upsilon\,\mathrm{sign}(x),\quad
  \upsilon_y = s\,\sin(\pi x),\quad\text{and}\quad
  \upsilon_z = 0.
\end{equation}
In Fig.~\ref{fig:shear:shock}, we present results for various SPH simulations
as well as a grid-code run for $s=100\delta\upsilon=100c$.
The M\&M method produces a large viscosity due to the shear-induced errors
in $\divv$, leading to spurious results. Using the Balsara limiter
with either M\&M or Standard SPH gives in much better results, though the
shock is clearly over-smoothed. The new scheme is able to limit the viscosity
to the correct level, allowing good capturing of the shock and retaining
particle order in the post-shock region.

Note that this is a difficult test for any SPH implementation: without
viscosity reduction (as in standard SPH) the shear flow is strongly damped,
while viscosity reduction schemes (M\&M as well as ours) suffer from the
problem of shear-induced errors. These potentially result in too much viscosity
and over-smoothing of the shock. Our limiter was able to control this problem,
but for yet larger ratios $s/\delta\upsilon$ of shear to shock amplitude
this problem becomes too difficult for any SPH implementation.

\begin{figure*}
  \centerline{
    \resizebox{155mm}{!}{\includegraphics{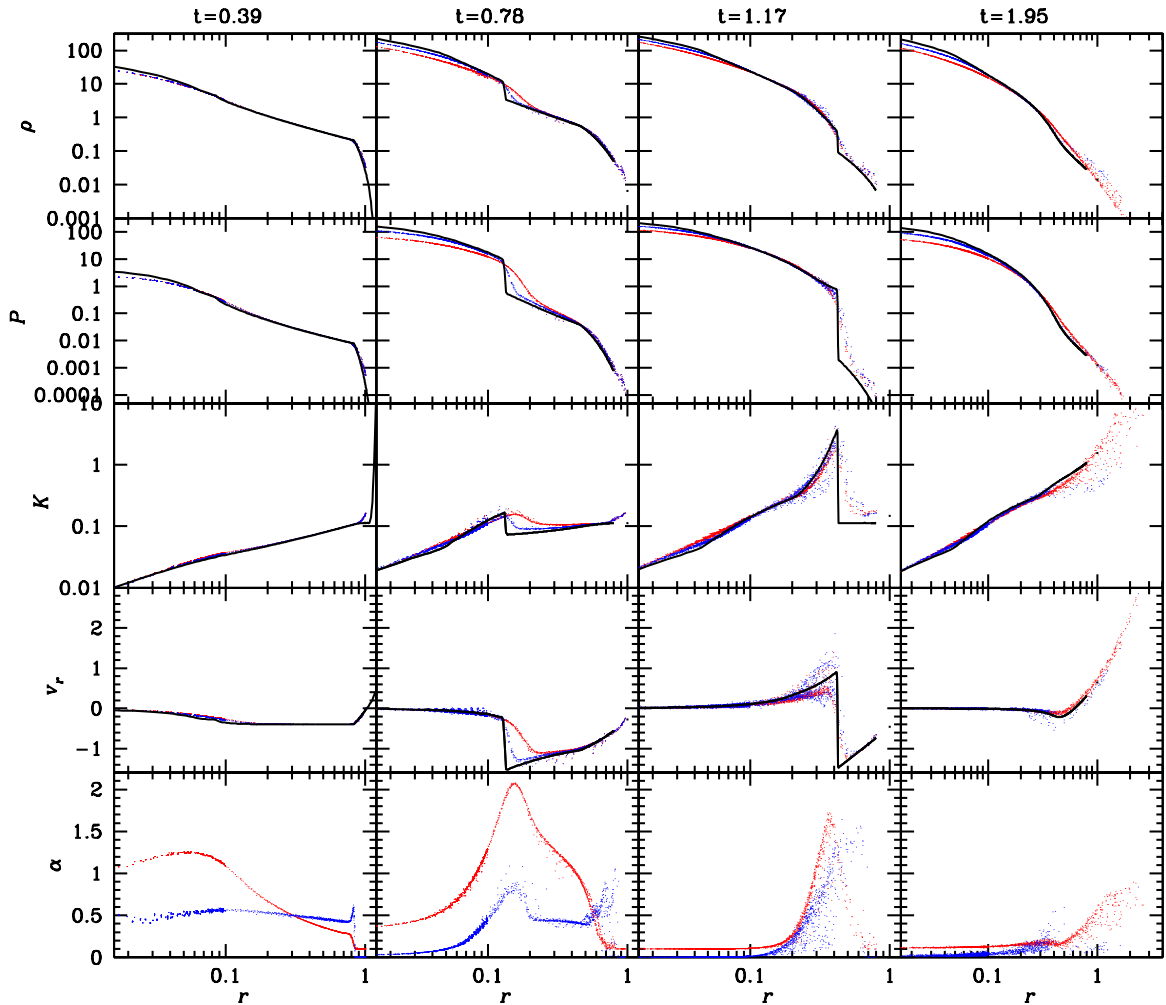}}
  }
  \caption{\label{fig:evrard} The Evrard test (see text for the initial setup):
    shown are various physical quantities ($K=P\rho^{-\gamma}$ is the entropy
    function) and $\alpha$ at different times for SPH simulations with $N=10^5$
    particles using our new viscosity scheme (blue) or the original M\&M method
    (red). Also shown (black) are the results from 1D PPM calculation
    \citep{SteinmetzMueller1993}. Not every particle is plotted.}
\end{figure*}
\subsection{Evrard Test}
\label{sec:test:evrard}
In this test the inward gravitational pull of a gas cloud exceeds its outward
pressure force causing the cloud to collapse under its own self-gravity. The
initial conditions consist of a gas sphere with density profile
\citep{Evrard1988}
\begin{equation}
  \rho(r) = \frac{M}{2\pi R^2}\frac{1}{r}
\end{equation}
for $r<R$ and $\rho=0$ for $r>R$. Initially the gas is at rest and has
constant specific internal energy $u=0.05\,GM/R$, which corresponds to a
virial ratio $-2U/W=0.075\ll1$. The initial gravitational inward pull is the
same at each radius, while the pressure forces decline outwards, leading to
collapse and, as a consequence, formation of a shock, which steepens and
evolves into a strong shock propagating outwards as more incoming material
joins the jam. Even though the problem is initially spherically symmetric, the
SPH realisation of initial conditions cannot be exactly spherically symmetric
and the system may well evolve away from sphericity, for instance driven by
dynamical instabilities.

We use a unit system such that $G{\,=\,}R{\,=\,}M\,{=\,}1$ and represent the
cloud by 100280 SPH particles, initially placed on a face-centred-cubic grid
which is then radially stretched to match the density. Fig.~\ref{fig:evrard}
compares the simulation results for our method, the original M\&M method, and
a 1D calculation by \cite{SteinmetzMueller1993} using the piece-wise parabolic
method (PPM).

At early times ($t=0.39$, left column) the results from all three methods
match very well, but the M\&M scheme already shows a large viscosity. At later
times a shock forms (at $r\approx0.13$ by $t=0.78$), which moves outwards
until it reaches the end of the sphere, when a significant fraction of the gas
still has outwards velocities (by $t=1.95$). The most obvious difference
between the two SPH schemes is the amount of (artificial) dissipation: the
M\&M method is much more viscous, resulting in significant over-smoothing of
the shock front by $t=0.78$ accompanied by unphysical pre-shock heating as
visible in the entropy ($K$) profile. Our new scheme agrees better with the 1D
calculation, in particular in the inner (post-shock) regions. Note that with
our new method $\alpha$ peaks well before the shock arrives (at $t=1.17$),
while for the M\&M method the peak in $\alpha$ appears actually slightly after
the shock.

We found this a valuable test as early versions of our scheme tended to be far
too viscous, while our final version passes this test ahead of the M\&M
switch. Standard SPH (not shown in the figure) shows similar results, though
the shock at $t=0.78$ appears less smoothed than with the M\&M method but more
smoothed than with the new scheme.

\section{Summary}
\label{sec:summ}
Any hydrodynamical numerical method requires some form of artificial viscosity
in order to resolve shocks \citep[in grid methods, artificial viscosity is
  implicit in the Riemann solver,][]{Monaghan1997}. In grid codes, such as
\textsf{Ramses} \citep{Teyssier2002}, interpolation methods are employed to
effectively suppress artificial viscosity away from shocks. Most SPH
simulations to this date hardly use such precautions and, as a consequence,
adiabatic oscillations and shear-flows are damped. Note that this
affects state-of-the-art simulations of, e.g.\ galaxy formation, which usually
only employ \citeauthor{Balsara1995}'s (\citeyear{Balsara1995}) rather
inefficient method to reduce some adverse effects of artificial viscosity on
rotation discs.

The method of \cite{MorrisMonaghan1997}, which reduces the default amount of
artificial viscosity by an order of magnitude compared to standard SPH
practice, has only recently been recognised as advantageous. In this method,
explained in detail in \S\ref{sec:AV:MM}, individual artificial viscosities
$\alpha_i$ are adapted by integrating a differential equation. Though
constituting a major improvement, this method remains unsatisfactory, because
it still damps adiabatic oscillations and over-smoothes weak shocks, as we
argued in \S\ref{sec:AV:novel} and demonstrated in \S\ref{sec:test:suppress}.

In \S\ref{sec:AV:novel}, we present a novel method, which improves upon that of
\citeauthor{MorrisMonaghan1997} in four important ways.
\begin{itemize}
\item We set $\alpha_{\min}=0$ enabling $\alpha_i\to0$ away from shocks and
  effectively modelling the fluid as inviscid.
\item We use $\Dtdivv\equiv\mathrm{d}(\divv)/\mathrm{d}t<0$ rather than
  $\divv<0$ as shock indicator. This distinguishes pre-shock from post-shock
  regions (where $\Dtdivv>0$ but $\divv<0$) and discriminates much better
  between converging flows and weak shocks.
\item We set $\alpha_i$ directly to an appropriate local value
  $\alpha_{\mathrm{loc}}$, instead of growing it by integrating a differential
  equation.
\item We use an improved estimator for $\divv$ and $\Dtdivv$ and
  employ a limiter to avoid viscosity driven by shear-induced errors.
\end{itemize}
Together these novelties result in a significantly improved artificial
viscosity method, in particular the viscosity is increased to an appropriate
level well before an incoming shock. The implementation details, i.e.\
the precise way of setting $\alpha_{\mathrm{loc}}$ from $\Dtdivv$ and the exact
form of the limiter, may well be subject to improvements. Any reader who
considers modifying these details is advised to consider the behaviour
of the resulting method for a test suite comprising noise
suppression as well as shear and strong shocks, for example the tests
of Figures~\ref{fig:noise}, \ref{fig:ring}, and \ref{fig:shear:shock}.

For static equilibria $\divv=0$ and $\dot{\B{\upsilon}}=0$, and our new shock
indicator (as well as the M\&M shock indicator) are only triggered by velocity
noise. As long as particle order is maintained, such noise triggers only
negligible amounts of viscosity, unlike the situation with the M\&M method,
whose minimum viscosity $\alpha_{\min}=0.1$ is often sufficient to affect the
simulations (as demonstrated in \S\ref{sec:test:suppress}). Nonetheless, the
noise-induced viscosity is sufficient to suppress particle disorder, as
demonstrated in \S\ref{sec:AV:amin}.

For dynamic equilibria $\divv=0$ (and $\Dtdivv=0$) but
$\dot{\B{\upsilon}}\neq0$. However, in multi-dimensional flows strong shear
induces false detections of $\divv$ (and $\Dtdivv$), even with best possible
particle order, for reasons explained in Appendix~\ref{app:divv:fail}. In
simulations of differentially rotating discs, this problem strongly affects
the M\&M method (even when using the Balsara switch). We avoid this problem by
applying a limiter (see \S\ref{sec:shear}) as well as using improved
estimators for $\divv$ and $\Dtdivv$, see Appendix~\ref{app:divv:good} for
details. (Alternatively, if no strong shear flows are present, the standard
estimators should suffice, though still in conjunction with a limiter using
$|\curlv|$ as a proxy for the shear amplitude.)

These improved estimators also provide the full velocity and acceleration
gradient matrices for each particle (and increase the computational costs by
${\sim\,}30\%$). The knowledge of the velocity gradient matrix $\B{\s{V}}$ and
its traceless symmetric part, the shear $\B{\s{S}}$, is also useful for
implementing physical viscosity
\begin{equation}\label{eq:PV}
  \rho\,\dot{\B{\upsilon}} = \B{\nabla}\cdot\left[ \eta\,\B{\s{S}} +
    \zeta\,\mathrm{tr}(\B{\s{V}})\right]
\end{equation}
(with $\eta$ and $\zeta$ the shear and bulk viscosity coefficients) in SPH.

In sections \ref{sec:AV:amin}, \ref{sec:test:suppress}, and
\ref{sec:test:shock}, we demonstrate convincingly that our technique 
successfully deals with the following four situations.
\begin{description}
\item[\textbf{Shocks}] are resolved at least as well, if not better,
  than with any previous technique;
\item[\textbf{adiabatic oscillations},] such as sound waves or stellar
  pulsations, remain undamped over many periods, which was not possible with
  any previous SPH implementation;
\item[\textbf{strong shear flows},] such as in accretion discs, are modelled
  virtually inviscid, while shearing shocks are well resolved without being
  over-smoothed;
\item[\textbf{particle disorder}] is suppressed at least as well as with 
  the M\&M method.
\end{description}
In particular, in the regime of convergent flows and weak shocks our new
method is far superior to any previous scheme, which all required a
significant increase in resolution just to suppress adverse effects of
artificial viscosity.

\subsection*{Acknowledgements}
Research in theoretical astrophysics at Leicester is supported by a STFC
rolling grant. We thank all members of the Leicester Theoretical Astrophysics
group, in particular Graham Wynn and Justin Read, for helpful discussions,
Seung-Hoon Cha for contributing the GPH data to Fig.~\ref{fig:1D_sound},
Matthias Steinmetz for providing the PPM data of the Evrard test for
Fig.~\ref{fig:evrard}, and the anonymous referee for helpful comments and
enhanced scrutiny.


\appendix
\section{Details of the SPH scheme} \label{sec:scheme}
For completeness, we give here a brief description of our SPH method, which is
largely similar to previous methods, but may differ in some details.

\subsection{Density and adaptive smoothing lengths}
\label{sec:adapt}
Let $\nu$ denote the number of spatial dimensions, then we adapt the
individual smoothing lengths $h_i$ such that $h_i^\nu\,\hat{\rho}_i=M_h$ with
$M_h\equiv m N_h/V_\nu$ a global constant, defined in terms of the number
$N_h$ of neighbours, the mass $m$ of each SPH particle, and the volume $V_\nu
$ of the unit sphere. In this work, we use $N_h=5,\,13,\,$ and 40 for
$\nu=1,\,2,$ and 3 dimensions, respectively. Inserting the density estimator
(\ref{eq:est:rho}), we find
\begin{equation} 
  \label{eq:Mh}
  h_i^\nu\,\hat{\rho}_i = \sum_j m_j\,w(r_{i\!j})
  \quad
  \text{with}
  \quad r_{i\!j}\equiv|\B{x}_{i\!j}|/h_i,
\end{equation}
where we have re-written the SPH kernel as $W(|\B{x}_{i\!j}|,h_i)=h_i^{-\nu}
w(r_{i\!j})$ with the dimensionless 
function $w(r)$. For this
work, we employ the usual cubic spline kernel \citep{MonaghanLattanzio1985}
\begin{equation}
  w(r) = \binom{\nu+3}{3}\,\frac{1}{V_\nu(2-2^{-\nu})} \times \left\{
    \begin{array}{ll}
      1-6r^2(1-r)\;\; & r\le1/2, \\[0.5ex]
      2(1-r)^3    & 1/2<r<1, \\[0.5ex]
      0           & \mathrm{otherwise}.
    \end{array}\right.
\end{equation}
At each time step, the $h_i$ are adjusted by performing one Newton-Raphson
step in $\log h$-$\log(h^\nu\hat{\rho})$ space, i.e.\
\begin{equation} 
  \label{eq:h:adapt}
  h_i \leftarrow h_i \left(\frac{M_h}{h_i^\nu\hat{\rho}_i}\right)^{f_i/\nu}
\end{equation}
with a factor of order unity
\begin{equation} 
  \label{eq:f}
  f_i = - \nu
  \frac{\sum_j m_j\,w_{i\!j}}
       {\sum_j m_j\,r_{i\!j}^2\,\tilde{w}_{i\!j}},
\end{equation}
where $w_{i\!j}\equiv w(r_{i\!j})$ and $\tilde{w}(r)\equiv
w^\prime(r)\!/r$. This method converges extremely well, except when $h_i$ was
much too small. In this case, faster convergence can be achieved by
subtracting the self-contribution (which does not depend on $h_i$). Thus,
whenever $h_i^\nu\hat{\rho}_i<M_h$ we use instead of (\ref{eq:h:adapt})
\begin{equation} \label{eq:h:adapt:better}
  h_i \leftarrow h_i \left(\frac{M_h-m_i\,w(0)}
  {h_i^\nu\hat{\rho}_i-m_i\,w(0)}\right)^{\tilde{f}_i/\nu}
  \;\text{with}\;
  \tilde{f}_i = - \nu
  \frac{\sum_{j\ne i} m_i\,w_{i\!j}}
       {\sum_{j\ne i} m_j\,r_{i\!j}^2\,\tilde{w}_{i\!j}}.
\end{equation}
The time derivatives $\dot{h}_i$ are obtained by demanding
$\mathrm{d}(h_i^\nu\hat{\rho}_i)/\mathrm{d}t=0$:
\begin{equation} 
  \label{eq:hdot}
  \frac{\dot{h}_i}{h_i}=
  \frac{\sum_j m_j\,\B{\upsilon}_{i\!j}\cdot\B{x}_{i\!j}\,\tilde{w}_{i\!j}}
       {\sum_j m_j\,\B{x}_{i\!j}^2\,\tilde{w}_{i\!j}}.
\end{equation}
\subsection{Pressure forces} \label{app:pressure}
We use SPH equations of motion derived from the simple SPH Lagrangian
$\mathcal{L}=\sum_km_k(\frac{1}{2}\dot{\B{x}}^2_k - u_k)$. Together with the
relation\footnote{Alternatively, for an ideal-gas equation of state one may
  replace $u$ in the Lagrangian with $u=K\hat{\rho}^{\gamma-1}/(\gamma-1)$ and
  consider the entropy function $K=P\hat{\rho}^{-\gamma}$ to be constant
  \citep{SpringelHernquist2002}.}  $\mathrm{d}u/\mathrm{d}\rho=P/\rho^2$,
this gives
\begin{equation} 
\label{eq:vdot}
  \dot{\B{\upsilon}}_i 
  = -\frac{1}{m_i} \frac{\partial\mathcal{L}}{\partial\B{x}_i}
  = -\sum_i m_j\,\B{x}_{i\!j}
  \left(
  \frac{P_if_i}{\hat{\rho}_i^2h_i^{\nu+2}} \tilde{w}_{i\!j} +
  \frac{P_jf_j}{\hat{\rho}_j^2h_j^{\nu+2}} \tilde{w}_{ji}
  \right),
\end{equation}
where the factors $f_i$ and $f_j$ (equation \ref{eq:f}) arise from the fact
that the derivatives $\partial\hat{\rho}_k/\partial\B{x}_i$ have to be taken
at fixed $h_k^\nu\hat{\rho}_k$. The work done by these pressure forces has to
be balanced by
\begin{equation} 
  \label{eq:udot}
  \dot{u_i}
  = -\nu \frac{P_i\,\dot{h}_i}{\hat{\rho}_i\,h_i}
  = \frac{P_i\,f_i}{\hat{\rho}_i^2\,h_i^{\nu+2}}
  \sum_j m_j\,\B{\upsilon}_{i\!j}\cdot\B{x}_{i\!j}\,\tilde{w}_{i\!j}.
\end{equation}

\subsection{Artificial viscosity}
For the artificial viscosity drag and heating we actually use
\begin{eqnarray}
  \label{eq:dv:av}
  (\dot{\B{\upsilon}}_i)_{\mathrm{AV}} &=-& \sum_j m_j\,\B{x}_{i\!j}\,
  \frac{\tilde{\Pi}_{i\!j}}{2}
  \left(
  \frac{\alpha_if_i}{\hat{\rho}_ih_i^{\nu+2}} \tilde{w}_{i\!j} +
  \frac{\alpha_jf_j}{\hat{\rho}_jh_j^{\nu+2}} \tilde{w}_{ji}
  \right) \\
  \label{eq:du:av}
  (\dot{u}_i)_{\mathrm{AV}} &=\phantom{-}&
  \sum_j m_j\;\B{\upsilon}_{i\!j}\cdot\B{x}_{i\!j}\,
  \frac{\tilde{\Pi}_{i\!j}}{2} \frac{\alpha_if_i}{\hat{\rho}_ih_i^{\nu+2}}
  \tilde{w}_{i\!j}
\end{eqnarray}
with $\tilde{\Pi}_{i\!j}=-\mu_{i\!j}(\bar{c}_{i\!j}-b\mu_{i\!j})$, where
\begin{equation}
  \mu_{i\!j} =
  \begin{cases}
    \displaystyle
    \frac{2\,\B{\upsilon}_{i\!j}\cdot\B{x}_{i\!j}}
         {(r_{i\!j}^2+r_{ji}^2)\bar{h}_{i\!j}}
    & \text{for} \quad \B{\upsilon}_{i\!j}\cdot\B{x}_{i\!j} < 0, \\
    0 & \text{otherwise};
  \end{cases}
\end{equation}
while the parameter $b$ has the meaning of $\beta/\alpha$ for traditional
SPH. Note that equations (\ref{eq:dv:av}) and (\ref{eq:dv}) differ only by
$\mathcal{O}(\bar{h}_{i\!j}^2)$. The difference between equations
(\ref{eq:du:av}) and (\ref{eq:du}) is more pronounced since, similarly to
equation (\ref{eq:udot}), we do not symmetrise the contributions w.r.t.\ $i$
and $j$.

\subsection{Time Integration}
Our scheme employs a kick-drift-kick leap-frog time integrator, which is
second-order accurate.  With this scheme, a full (global) time step of size
$\delta t$ consists of the following sub-steps (`$\leftarrow$' means `is
replaced by').
\begin{description}
\item[\textbf{initial kick}] Compute $\B{\upsilon}_i$ and $u_i$
  at half step
  \begin{equation} 
    \label{eq:kick}
    \begin{array}{rcl}
      \tilde{\B{\upsilon}}_i &=& \B{\upsilon}_i + \tfrac{1}{2}\,\delta
      t\,\dot{\B{\upsilon}}_i, \\[0.5ex]
      \tilde{u}_i &=& u_i + \tfrac{1}{2}\,\delta
      t\,\dot{u}_i.
    \end{array}
  \end{equation}
\item[\textbf{full drift}] Advance $t$ and $\B{x}_i$ by a full step:
  \begin{equation} \label{eq:drift}
    \begin{array}{rcl}
      t       &\leftarrow& t + \delta t \\
      \B{x}_i &\leftarrow& \B{x}_i + \delta t\,\tilde{\B{\upsilon}}_i.
    \end{array}
  \end{equation}
\item[\textbf{prediction}] Predict $\B{\upsilon}_i$, $u_i$, and $h_i$ at full
  step:
  \begin{equation} \label{eq:pred}
    \begin{array}{rcl}
      \B{\upsilon}_i &\leftarrow& \B{\upsilon}_i + 
      \delta t\,\dot{\B{\upsilon}}_i, \\
      u_i     &\leftarrow& u_i\exp(\delta t\,\dot{u}_i/u_i), \\
      h_i     &\leftarrow& h_i\exp(\delta t\,\dot{h}_i/h_i).
    \end{array}
  \end{equation}
\item[\textbf{sweep 0}] Compute $h_i^\nu\hat{\rho}_i$ and $f_i$ (equations
  \ref{eq:Mh} and \ref{eq:f}).
\item[\textbf{adapt}] Adjust $h_i$ (equation \ref{eq:h:adapt} or
  \ref{eq:h:adapt:better}).
\item[\textbf{sweep 1}] Compute $\hat{\rho}_i$, $f_i$, and $\dot{h}_i$
  (eqs.~\ref{eq:Mh}, \ref{eq:f}, and \ref{eq:hdot}) as well as $\divv_i$,
  $\Dtdivv_i$, and $R_i$ (eqs.~\ref{eq:divv:new}, \ref{eq:Dtdivv}, and
  \ref{eq:R}, using $\dot{\B{\upsilon}}$ and $\divv$ from the previous time
    step).
\item[\textbf{between sweeps}] Obtain $P_i$ and $c_i$ from $\hat{\rho}_i$ and
  $u_i$ via the equation of state, and adapt $\alpha_i$ via
  (using eqs.~\ref{eq:tau} and \ref{eq:a:loc})
  \begin{equation} 
    \label{eq:int:av}
    \alpha_i \leftarrow
    \begin{cases}
      \alpha_{\mathrm{loc}} & \text{if $\alpha_i<\alpha_{\mathrm{loc}}$}, \\
      \alpha_{\mathrm{loc}} + (\alpha_i-\alpha_{\mathrm{loc}})
      \exp(-\delta t/\tau_i) & \text{otherwise}.
    \end{cases}
  \end{equation}
\item[\textbf{sweep 2}] Compute $\dot{\B{\upsilon}}_i$ (eqs.~\ref{eq:vdot} and
  \ref{eq:dv:av} plus gravitational forces) and $\dot{u}_i$
  (eqs.~\ref{eq:udot} and \ref{eq:du:av} plus external heating or cooling).
\item[\textbf{final kick}] Set $\B{\upsilon}_i$ and $u_i$  at full step: 
  \begin{equation} \label{eq:final}
    \begin{array}{rcl}
      \B{\upsilon}_i &=& \tilde{\B{\upsilon}}_i + \tfrac{1}{2}\,\delta t\,
      \dot{\B{\upsilon}}_i, \\[1ex]
      u_i     &=& \tilde{u}_i + \tfrac{1}{2}\,\delta t\,\dot{u}_i.
    \end{array}
  \end{equation}
\end{description}
In the initial kick and prediction steps, the time derivatives are known from
the previous time step (in case of the very first time step, they need to be
precomputed). Note that the quantities predicted in (\ref{eq:pred}) enter the
final $\B{\upsilon}_i$ and $u_i$ only indirectly via the computation of the
time derivatives.

We use an oct-tree, generated just before sweep 0, to find all interacting
particle pairs, which are then remembered in an interaction list, whereby
allowing for the fact that $h_i$ may grow slightly during adjustment (just
after sweep 0). Utilising this interaction list in sweeps 1 and 2 is much
faster than further tree walks. The same oct-tree is also used in computing
gravitational forces, as outlined by \cite{Dehnen2002}.

Our scheme can also be implemented with adaptive individual time steps
organised in a hierarchical block-step scheme, though we have not used this in
the tests presented in this study.

\section{\boldmath Estimating $\divv$ and $\Dtdivv$}
\label{app:divv}
\subsection{Failure of the standard SPH estimator for $\divv$}
\label{app:divv:fail}
Our constraint that $h_i^\nu\hat{\rho}_i$ be constant (see \S\ref{sec:adapt})
implies $\dot{\hat{\rho}}_i/\hat{\rho}_i=-\nu\dot{h}_i/h_i$. Together with the
continuity equation $\dot{\rho}+\rho\,\divv=0$ and equation~(\ref{eq:hdot})
this yields the simple velocity-divergence estimate
\begin{equation} \label{eq:divv:std}
  \widehat{\divv}_i = \nu \frac{\sum_j
      m_j\,\B{\upsilon}_{i\!j}\cdot\B{x}_{i\!j}\,\tilde{w}_{i\!j}} {\sum_j
      m_j\,\B{x}_{i\!j}^2\,\tilde{w}_{i\!j}}.
\end{equation}
While this estimate satisfies the continuity equation for the SPH density
estimate $\hat{\rho}_i$, it is not necessarily accurate. To see this, consider
the matrix ($\otimes$ denotes the outer or dyadic vector product)
\begin{equation} \label{eq:D} \textstyle
  \B{\s{D}}_i = \sum_j \B{\upsilon}_{i\!j}\otimes \B{x}_{i\!j}\,\bar{w}_{i\!j}
\end{equation}
with $\bar{w}_{i\!j}$ some weighting factor. Assuming a smooth velocity field,
we may replace $\B{\upsilon}_{i\!j}$ in equation~(\ref{eq:D}) with its Taylor
expansion $\B{\upsilon}_{i\!j} = \B{\s{V}}_i\cdot\B{x}_{i\!j} +
\mathcal{O}(|\B{x}_{i\!j}|^2)$, where $\B{\s{V}}_{\!i}\equiv
\B{\nabla}\otimes\B{\upsilon}|_{\B{x}_i}$ is the gradient of $\B{\upsilon}$ at
position $\B{x}_i$, and obtain
\begin{equation} \label{eq:DVT}
  \B{\s{D}}_i = \B{\s{V}}_{\!i}{\cdot}\B{\s{T}}_{\!i} + \mathrm{h.o.t.}
\end{equation}
with the symmetric matrix
\begin{equation} \label{eq:T}\textstyle
  \B{\s{T}}_{\!i} = \sum_j
  \B{x}_{i\!j}\otimes\B{x}_{i\!j}\,\bar{w}_{i\!j}.
\end{equation}
Comparing (\ref{eq:D}) and (\ref{eq:T}) to the simple estimator
(\ref{eq:divv:std}), we see that the latter corresponds to (conveniently
dropping the index $i$) $\widehat{\divv}= \nu\, \mathrm{tr} (\B{\s{D}})/
\mathrm{tr}(\B{\s{T}})$ and the weights
$\bar{w}_{i\!j}=m_{\!j}\tilde{w}_{i\!j}$. If we split $\B{\s{V}}$ into its
isotropic part (divergence), the symmetric traceless part $\B{\s{S}}$ (shear),
and the antisymmetric part $\B{\s{R}}$ (vorticity),
\begin{equation} \label{eq:V=d+S+R}
  \B{\s{V}} = \nu^{-1}\,\divv\, \B{\s{I}} + \B{\s{S}} + \B{\s{R}},
\end{equation}
and insert it into~(\ref{eq:DVT}), we find for the simple
estimator (\ref{eq:divv:std})
\begin{equation}
  \widehat{\divv} = \divv
  + \nu\,\mathrm{tr}(\B{\s{S}}{\cdot}\tilde{\B{\s{T}}})/\mathrm{tr}(\B{\s{T}})
  + \mathrm{h.o.t.}
\end{equation}
where $\tilde{\B{\s{T}}}$ denotes the anisotropic (traceless) part of
$\B{\s{T}}$. Thus, the simple estimator~(\ref{eq:divv:std}) contains an
$\mathcal{O}(h^0)$ error term, which originates from anisotropy of $\B{\s{T}}$
in conjunction with velocity shear (owing to the symmetry of $\B{\s{T}}$ the
vorticity is harmless). For perfectly symmetric particle distributions
$\tilde{\B{\s{T}}}=0$, but in general $\tilde{\B{\s{T}}}\neq0$ such that in
the presence of strong shear even a small residual $\tilde{\B{\s{T}}}$ results
in a failure of the simple estimator (\ref{eq:divv:std}). This typically
happens in differentially rotating discs, where (i) the velocity field is
divergent-free but contains shear and (ii) even in the absence of noise
$\tilde{\B{\s{T}}}\neq0$ owing to the shearing particle distribution.

\subsection{\boldmath A more accurate $\divv$ estimator}
\label{app:divv:good}
From equation~(\ref{eq:DVT}), we can also estimate
\begin{equation} \label{eq:V:est}
  \hat{\B{\s{V}}}_{\!i} = \B{\s{D}}_i{\cdot}\B{\s{T}}_{\!i}^{-1},
\end{equation}
which allows an improved divergence estimator 
\begin{equation} \label{eq:divv:new}
  \widehat{\divv}_i =
  \mathrm{tr}\left(\B{\s{D}}_i{\cdot}\B{\s{T}}_{\!i}^{-1}\right).
\end{equation}
In order to assess the error of this estimator, let us expand the flow to
second order, replacing equation~(\ref{eq:DVT}) with (dropping the index $i$
and using suffix instead of matrix notation)
\begin{equation} \label{eq:D:err}
  \s{D}_{\alpha\beta} = \upsilon_{\alpha,\gamma}\,\s{T}_{\gamma\beta}
  - \tfrac{1}{2} \upsilon_{\alpha,\gamma\delta}\,\s{U}_{\gamma\delta\beta}
  + \mathrm{h.o.t.}
\end{equation}
with the symmetric tensor $\B{\s{U}}_i=\sum_j \B{x}_{i\!j} \otimes\B{x}_{i\!j}
\otimes\B{x}_{i\!j} \,\bar{w}_{i\!j}$. Inserting this into~(\ref{eq:V:est}) we
find
\begin{equation} \label{eq:V:err}
  \hat{\s{V}}_{\alpha\beta} = \upsilon_{\alpha,\beta}
  - \tfrac{1}{2} \upsilon_{\alpha,\gamma\delta}\,\s{U}_{\gamma\delta\eta}
  \s{T}^{-1}_{\eta\beta} + \mathrm{h.o.t.}.
\end{equation}
Thus, while this estimator avoids an $\mathcal{O}(h^0)$ error, we still have
an $\mathcal{O}(h^1)$ error term (since $\B{\s{U}}$ is one order higher in $h$
than $\B{\s{T}}$). We can \emph{reduce} the $\mathcal{O}(h^1)$ error by a
careful choice of the weights $\bar{w}_{i\!j}$. If, for instance,
$\bar{w}_{i\!j}=m_{\!j}\tilde{w}_{i\!j}/\hat{\rho}_{\!j}$ then $\B{\s{U}}\to0$
to leading order in the \emph{continuum limit} by virtue of the isotropy of
the kernel. This limit, which is commonly used to assess SPH estimators,
replaces $\sum_j m_j$ with $\int\rho(\B{x}_{\!j})\, \mathrm{d}\B{x}_{\!j}$
under the assumption of a smooth density without particle noise\footnote{Under
  these conditions also $\tilde{\B{\s{T}}}$, which causes the
  $\mathcal{O}(h^0)$ error term in the simple $\divv$ estimator,
  vanishes.}. As these conditions are hardly ever truly satisfied, we can only
reduce but not eliminate the $\mathcal{O}(h^1)$ error term---as we do not even
try to avoid the $\mathcal{O}(h^2)$ error (hidden in `h.o.t.' above), such
a reduction should be okay in most cases.

\subsection{\boldmath Estimating $\Dtdivv$}
\label{app:divv:dot}
We can estimate $\Dtdivv$ either from the change in the estimated $\divv$ over
the last time step or as the trace of $\dot{\B{\s{V}}}$, the total time
derivative of $\B{\s{V}}$. Since (with $\B{\s{A}}\equiv
\B{\nabla}\otimes\dot{\B{\upsilon}}$ the gradient of the acceleration)
\begin{equation} \label{eq:Vdot=A-V^2}
  \dot{\B{\s{V}}} = \B{\s{A}} - \B{\s{V}}^2
\end{equation}
(a good exercise for your undergraduate students), we can estimate
\begin{equation} \label{eq:Dtdivv}
  \widehat{\Dtdivv}_i = \mathrm{tr}\Big(\hat{\B{\s{A}}}_i -
  \hat{\B{\s{V}}}_i^2\Big).
\end{equation}
Here, the estimate $\hat{\B{\s{A}}}_i$ is obtained from the accelerations at
the previous time step in a way analogous to estimating
$\hat{\B{\s{V}}}_{\!i}$, in particular we need to compute the matrix
$\B{\s{T}}_{\!i}$ and its inverse only once. The lowest-order error in this
estimate again is $\mathcal{O}(h^1)\propto\B{\s{U}}_i$, such that reducing
$\B{\s{U}}_i$ by careful choice of the weights remains a good idea.

Note that, by virtue of equation~(\ref{eq:Vdot=A-V^2}), we could estimate
$\Dtdivv$ also as $\diva-\mathrm{tr}(\B{\s{V}}^2)$ with the acceleration
divergence $\diva$ estimated using the standard divergence estimator, in the
hope that its $\mathcal{O}(h^0)$ error term is small since the acceleration
is hardly sheared.

\label{lastpage}
\end{document}